\documentclass[3p]{elsarticle}
\usepackage{amsmath,amsfonts,amssymb}
\usepackage{algorithmic}
\usepackage{algorithm}
\usepackage{graphicx}
\usepackage[multiple]{footmisc}

\usepackage{array}
\usepackage[caption=false,labelfont=sf,textfont=sf]{subfig}
\usepackage{textcomp}
\usepackage{stfloats}
\usepackage{url}
\usepackage{verbatim}
\usepackage{graphicx,color,xcolor}
\usepackage{multirow}
\usepackage{booktabs}
\usepackage{fdsymbol}
\usepackage{colortbl}
\usepackage{hyperref}
\hypersetup{
    colorlinks=true,
    linkcolor=purple,
    filecolor=magenta,      
    urlcolor=purple,
}

\definecolor{colortDCF}{RGB}{0, 103, 181}
\definecolor{colorEER}{RGB}{210, 72, 24}

\newenvironment{partnum}[1]{\texttt{\#{#1}}}

\begin{document}

\begin{frontmatter}
\title{ASVspoof 5: Design, Collection and Validation of Resources for Spoofing, Deepfake, and Adversarial Attack Detection Using Crowdsourced Speech}

\author[nii]{Xin Wang}\ead{wangxin@nii.ac.jp}
\author[microsoft]{H\'ector Delgado}\ead{hector.delgado@microsoft.com}
\author[pindrop]{Hemlata Tak}\ead{Hemlata.Tak@pindrop.com}
\author[cmu]{Jee-weon Jung}\ead{jeeweonj@ieee.org (currently at Apple)}
\author[cmu]{Hye-jin Shim}\ead{hyejinsh@andrew.cmu.edu}
\author[eurecom]{Massimiliano Todisco}\ead{Massimiliano.Todisco@eurecom.fr}
\author[klass]{\\Ivan Kukanov}\ead{Ivan@kukanov.com}
\author[nii]{Xuechen Liu}\ead{xuecliu@nii.ac.jp}
\author[tcg]{Md Sahidullah}\ead{sahidullahmd@gmail.com}
\author[uef]{Tomi Kinnunen}\ead{tkinnu@cs.uef.fi}
\author[eurecom]{Nicholas Evans}\ead{evans@eurecom.fr} 
\author[polyu]{Kong Aik Lee}\ead{kong-aik.lee@polyu.edu.hk}
\author[nii]{\\Junichi Yamagishi}\ead{jyamagis@nii.ac.jp}
\author[snu]{Myeonghun Jeong}\ead{mhjeong@hi.snu.ac.kr}
\author[roc]{Ge Zhu}\ead{gzhu@adobe.com (currently at Adobe Research)}
\author[roc]{Yongyi Zang}\ead{yzang4@u.rochester.edu}
\author[roc]{You Zhang}\ead{yzh298@ur.rochester.edu}
\author[cmu]{Soumi Maiti}\ead{soumimaiti@meta.com (currently at Meta)}
\author[stuggart]{Florian Lux}\ead{florian.lux@ims.uni-stuttgart.de}
\author[fraunhofer]{Nicolas M\"{u}ller}\ead{nicolas.mueller@aisec.fraunhofer.de}
\author[sjtu]{Wangyou Zhang}\ead{zwyemrys@gmail.com}
\author[buffalo]{Chengzhe Sun}\ead{csun22@buffalo.edu}
\author[buffalo]{Shuwei Hou}\ead{shuweiho@buffalo.edu}
\author[buffalo]{Siwei Lyu}\ead{siweilyu@buffalo.edu}
\author[helsinki]{S\'ebastien Le Maguer}\ead{sebastien.lemaguer@helsinki.fi}
\author[tju]{Cheng Gong}\ead{gongchengcheng@tju.edu.cn}
\author[ustc]{Hanjie Guo}\ead{ghj2001@mail.ustc.edu.cn}
\author[ustc]{Liping Chen}\ead{lipchen@ustc.edu.cn}
\author[uef]{Vishwanath Singh}\ead{vishwanath.singh@uef.fi}

\address[nii]{National Institute of Informatics, 2-1-2 Hitotsubashi, Chiyoda-ku, Tokyo, Japan} 
\address[microsoft]{Microsoft, P.º Club Deportivo, 1, Edificio 1, 28223 Pozuelo de Alarcón, Madrid, Spain}
\address[pindrop]{Pindrop, 1115 Howell Mill Rd NW \#700, 30318, Atlanta GA, USA}
\address[cmu]{Carnegie Mellon University, 5000 Forbes Avenue, 15213, Pittsburgh, USA}
\address[eurecom]{EURECOM, Campus SophiaTech, 450 Route des Chappes, 06410 Biot, France}
\address[klass]{KLASS Engineering and Solutions, 30A Kallang Pl, \#11-03, 339213 Singapore}
\address[tcg]{Institute for Advancing Intelligence, TCG CREST, 700091, Kolkata, India}
\address[uef]{University of Eastern Finland, Joensuu campus, L\"{a}nsikatu 15, FI-80110 Joensuu, Finland}
\address[polyu]{The Hong Kong Polytechnic University, Kowloon, Hong Kong, China}
\address[snu]{Seoul National University, 1 Gwanak-ro, Gwanak-gu, 08826, Seoul, Republic of Korea}
\address[roc]{University of Rochester, 720 Computer Studies Building, 14627, Rochester, USA}
\address[stuggart]{University of Stuttgart, Pfaffenwaldring 5 b, 70569, Stuttgart, Germany}
\address[fraunhofer]{Fraunhofer AISEC, Lichtenbergstrasse 11, 85748 Garching, Germany}
\address[sjtu]{Shanghai Jiao Tong University, 200240, Shanghai, China}
\address[buffalo]{University at Buffalo, 14260 NY, Buffalo USA}
\address[helsinki]{University of Helsinki,  FI-00100, Helsinki, Finland}
\address[tju]{Tianjin University, No.135 Yaguan Road, 300350, Tianjin, China}
\address[ustc]{University of Science and Technology of China, No.96, JinZhai Road, 230026, Hefei, China}

\begin{abstract}
ASVspoof~5 is the fifth edition in a series of challenges which promote the study of speech spoofing and deepfake attacks as well as the design of detection solutions. 
We introduce the ASVspoof~5 database which is generated in a crowdsourced fashion from data collected 
in diverse acoustic conditions (cf.\ studio-quality data for earlier ASVspoof databases) and from 
$\sim$2,000 speakers (cf.\ $\sim$100 earlier).
The database contains attacks generated with 32 different algorithms, also crowdsourced, and optimised to varying degrees using new surrogate detection models.  Among them are attacks generated with a mix of legacy and contemporary text-to-speech synthesis and voice conversion models, in addition to adversarial attacks which are incorporated for the first time. 
ASVspoof~5 protocols 
comprise seven speaker-disjoint partitions. 
They include two distinct partitions for the training of different sets of attack models, two more for the development and evaluation of surrogate detection models, and then three additional partitions which comprise the ASVspoof~5 training, development and evaluation sets.
An auxiliary set of data collected from an additional 30k speakers can also be used to train speaker encoders for the implementation of attack algorithms.
Also described herein is an experimental validation of the new ASVspoof~5 database using a set of automatic speaker verification and spoof/deepfake baseline detectors. 
With the exception of protocols and tools for the generation of spoofed/deepfake speech, the resources described in this paper, already used by participants of the ASVspoof~5 challenge in 2024, are now all freely available to the community.
\end{abstract}

\end{frontmatter}

\section{Introduction}
\label{sec:intro}

The potential threats and risks related to the misuse of deepfakes---synthetic or manipulated media generated with the aid of deep learning---are well acknowledged both by various professional communities (e.g.\ security, forensics, biometrics) and government bodies, as well as by the general public. Having emerged as a novel type of information security threat which covers multiple domains---news, social media, and communication to name a few---it is more timely than ever to develop proactive defences against deepfakes. The focus of this article is specifically \emph{speech} deepfakes. Speech, as the primary means of human communication, carries not only the linguistic message but also \emph{paralinguistic} (beyond language) information.  This includes voice timbre in addition to other personally identifiable attributes, representations of which can be estimated from just a few seconds of speech and then used to infer the speaker identity using automatic speaker verification (ASV) technology. The voice timbre can then be transplanted into synthetic or manipulated speech to impersonate a specific, target individual.

The most widely studied solution to defend against the potentially harmful impact of speech deepfakes takes the form of \emph{detection} solutions which provide a means to distinguish between bona fide (genuine) and synthetic or manipulated (spoofed) speech. Their reliability is strongly dependent on the availability of realistic and representative \emph{data} with which a detection model is implemented and trained.  Among other factors which complicate the design of reliable detection solutions is the diversity in text-to-speech (TTS) synthesis and voice conversion (VC) technology now available (even to the layman) to generate or manipulate speech signals. Furthermore, the multitude of different model architectures and training algorithms continues to evolve at an astonishing pace, implying that detectors trained even only a few months ago have the potential to fail in the face of deepfakes generated using technologies which have emerged since.  Media distribution platforms alike continue to evolve, as do the encoding and lossy compression algorithms they use, all of which can introduce distortions which interfere with or perturb a detection system trained without identically treated or sufficiently similar data.

Born from a special session held at Interspeech in 2013, the efforts to develop detection solutions for speech data were initially spearheaded through the \textbf{ASVspoof initiative and challenge series}, founded over a decade ago~\cite{interspeechSpecialSession2013, Wu-ASVspoof2015,Wu2017-ASVspoof-IEEE-J-STSP, Kinnunen2017-assessing, asvspoof2019, yamagishiASVspoof2021Accelerating2021, wang2024asvspoof5}.  ASVspoof has evolved from a challenge in the detection of spoofing attacks implemented with now-legacy TTS and VC algorithms to encompass also the detection of deepfake attacks implemented with the most recent deepfake technology as well as their impact upon the reliability of ASV systems. The tracking of developments in the speech synthesis and voice conversion communities is key to the pursuit of reliable and robust detection solutions. Each challenge edition was hence accompanied by the collection and public release of an updated ASVspoof database.  The most recent challenge edition, ASVspoof~5, is no exception.  The design, collection and validation of such data is a non-trivial undertaking and requires coordinated and collaborative international effort; in this article we summarise the collective effort of 2.5 years of work conducted by members of 18 teams spanning three continents. 

Earlier ASVspoof databases were generated predominantly from the Voice Cloning Tool Kit (VCTK) database~\cite{yamagishiCSTRVCTKCorpus2019}. It contains studio-quality recordings of speech collected in a hemi-anechoic chamber and hence lacks distortions which might typify recordings collected in the wild, e.g.\ additive and convolutional noise, non-linear noise, coding and compression artefacts, narrow bandwidths, etc.  Evaluation results derived using such databases might not provide the most reliable estimates of performance which could be expected in practical scenarios.  Since spoof/deepfake attacks are generated using \emph{clean} training and speaker-specific adaptation data, the resulting high quality of synthesized or converted speech might inflate increases in the false alarm/acceptance rate for both ASV and spoof/deepfake detection systems, referred to as countermeasures (CMs); an adversary is unlikely to have the luxury of using studio-quality adaptation data collected from the victim.  For the same reason, the reliability of CMs, when trained with data of matched, high quality might also be exaggerated.

The adoption of a new source database hence offers an opportunity to provide the community with data which is more representative of that likely to be encountered in the wild.  This, in turn, should help provide more reliable estimates of the impact of attacks as well as CM performance.  It also offers an opportunity to provide data collected from a greater number of speakers as well as to assess detection performance when the quantity of TTS/VC training/adaptation data varies.  All earlier ASVspoof databases contain data collected from $\sim$100 speakers.\footnote{
The ASVspoof 2015~\cite{Wu-ASVspoof2015}, 2019 Logical Access~\cite{asvspoof2019database}, 2021 Logical Access and Deepfake databases~\cite{yamagishiASVspoof2021Accelerating2021} contain data collected from $\sim$100 speakers sourced from the VCTK database. 
The ASVspoof 2021 Deepfake database contains data collected from an additional 26 speakers sourced from the Voice Conversion Challenge 2018~\cite{Lorenzo-Trueba2018} and 2020~\cite{Yi2020} databases. The ASVspoof 2017~\cite{Kinnunen2017-assessing} and 2019 Physical Access databases, which contain data collected from 42 and 106 speakers respectively, focus on replay spoofing attacks.}
The low number of speakers is a constraint on not just the protocol design, but also on the potential to train speaker-independent CMs.  The adoption of a new source database also necessitates the design of new attack algorithms capable of generating spoofs/deepfakes using lower-quality training and adaptation data, as well as contemporary coding and compression techniques, specifically those developed since the last ASVspoof challenge held in 2021.  Last, we have been alerted to the presence of \emph{shortcut} artefacts~\cite{geirhos2020shortcut} which plagues some earlier ASVspoof databases~\cite{chettriDataset2020,muller21_asvspoof,shim23b_interspeech,zhang2023impact,liuASVspoof2021Spoofed2023}, namely artefacts which are semantically unrelated to the spoof/deepfake problem but which can nonetheless be utilised for detection.  The collection of a new database hence offers an opportunity to reduce such biases. 

\begin{figure*}[t!]
    \centering
     \includegraphics[width=\textwidth, trim=0 500 0 0, clip]{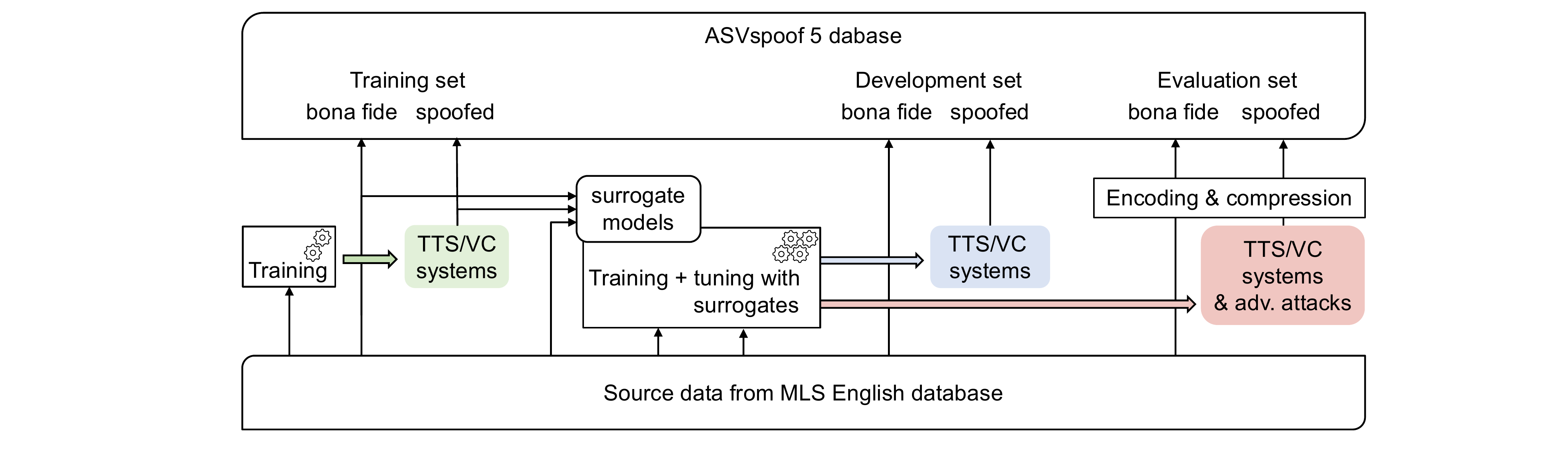}
     \caption{Overview of the ASVspoof~5 database. Bona fide utterances contained in the speaker-disjoint training, development, and evaluation sets are sourced directly from the MLS English database. Spoofed utterances in each of the three sets are generated using distinct TTS and VC attack algorithms and optionally combined with supplementary adversarial attacks. All attacks are trained using held-out training data again sourced from the MLS English database. Attacks in the development and evaluation sets are optionally tuned using surrogate ASV and CM models, both trained using bona fide and spoofed data contained within the ASVspoof 5 training set in addition to additional, reserved data.}
     \label{fig:overview}
\end{figure*}

As illustrated in Figure~\ref{fig:overview}, the new \textbf{ASVspoof~5 database}\footnote{ASVspoof~5 database link: \url{https://doi.org/10.5281/zenodo.14498691}} is generated from the English partition of the Multilingual Librispeech (MLS) database~\cite{pratap20_interspeech}.  It is itself sourced from LibriVox~\cite{kearns2014librivox}, a collection of free public domain audiobooks contributed by volunteers from around the world.\footnote{\url{https://librivox.org/}}  Having been collected in each contributor's own recording setting, the MLS source database contains far greater acoustic variation than the VCTK database.  The MLS database furthermore contains speech recordings collected from thousands of speakers.  This order-of-magnitude increase in the number of speakers (cf.\ previous ASVspoof databases) permits greater flexibility in database and protocol design, in particular so that attack algorithms can be tuned using a set of \emph{surrogate} CM and ASV systems (middle of Figure~\ref{fig:overview}) trained, developed and evaluated using held-out data.  Spoofing and deepfake attacks contained within the ASVspoof~5 training, development and evaluation sets (green, blue and red boxes in Figure~\ref{fig:overview}) are generated by a group of data contributors, all experts in TTS and VC, again using disjoint data.  Adversarial attacks~\cite{szegedy2013intriguing, goodfellow2014explaining} are introduced (evaluation set only, red box in Figure~\ref{fig:overview}) for the first time, as is the use of neural codecs and a new pipeline to tackle the biases of shortcut artefacts. The latter is used to reduce the mismatch between distributions of peak waveform amplitude, the duration of leading and trailing non-speech segments and whole utterance duration, for bona fide and spoofed/deepfake utterances.

We describe the design of the ASVspoof~5 database (\S~\ref{sec:database_overview}), the crowd-sourcing approach to the collection of spoofing/deepfake attacks (\S~\ref{sec:database_overview}-\ref{sec:attacks}), the pipeline to reduce shortcut artefacts (\S~\ref{sec:postprocess}), data visualisations to illustrate the similarities and differences between different attacks (\S~\ref{sec:visualize}), and an experimental validation using CM and ASV baselines (\S~\ref{sec:experiment}). A presentation of challenge results, beyond that already available in~\cite{wang2024asvspoof5}, is in preparation for later publication. 

\section{Database generation}
\label{sec:database_overview}

We describe the design of the ASVspoof~5 protocols and database and its generation from new source data.  The treatment is dense and, for this reason, readers who are interested only in using the ASVspoof~5 database for their own research in detection tasks (speakers or attacks) should consider reading at least Sections~\ref{subsec:mls} and~\ref{subsec:partition}.  These readers might then prefer to move directly either to Section~\ref{sec:attacks} or to Section~\ref{sec:visualize}.  Readers with a keen eye for %
protocol design, or those with an interest in using ASVspoof~5 protocols to generate their own attacks\footnote{For ethical reasons, attack generation protocols are not shared publicly. Those interested in the use of attack generation protocols to support further research in detection can request access to the generation protocols by sending an email to the \href{mailto:organisers@lists.asvspoof.org}{ASVspoof organisers}.} should consider reading this section in its entirety.  The focus throughout is upon source data and protocols.  A description of specific TTS/VC systems and adversarial attacks is provided in Section~\ref{sec:attacks}.

\subsection{Source database}
\label{subsec:mls}
Earlier ASVspoof databases were generated using primarily the VCTK~\cite{yamagishiCSTRVCTKCorpus2019} source database which contains utterances collected from approximately 100 speakers in a single, hemi-anechoic chamber.  This choice was made to ease the generation of spoofed speech yet, as a consequence, both bona fide and spoofed utterances are of higher quality than might be expected in practice.  Adversaries may also not be able to acquire training data of similar studio quality. The evaluation of detection performance in the face of non-studio-quality data calls for the adoption of an alternative source database.

The ASVspoof~5 database is generated using the MLS English database~\cite{pratap20_interspeech}.  
It contains data collected from approximately 2,400 female and 2,300 male speakers in diverse recording conditions (acoustic environments and devices). 
The MLS database is itself sourced from LibriVox~\cite{kearns2014librivox}, a collection of free, public domain audiobooks. Recordings are made by individual contributors in their own home or office and with their own recording devices.  The MLS database hence contains speech data captured in variable, non-studio-quality conditions.  Utterances contained within the MLS English database are of 10-to-20 seconds and originate from the segmentation of LibriVox audiobook recordings.

As illustrated in Figure~\ref{fig:ASVspoof5_protocol}, the MLS database is partitioned into 7 speaker-disjoint subsets.  They are constructed carefully according to a number of speaker/utterance criteria explained later in Section~\ref{subsec:spkutt}.  First, we describe the purpose of each partition, starting with those which comprise the ASVspoof~5 database, then a pair used for the training of TTS/VC systems and adversarial attacks, followed by a pair of auxiliary partitions related to the use of surrogate CM and ASV systems.

\subsection{ASVspoof~5 database partitions}
\label{subsec:partition}

\begin{figure*}[t!]
     \centering
     \includegraphics[width=0.99\textwidth, trim=0 0 0 0, clip]{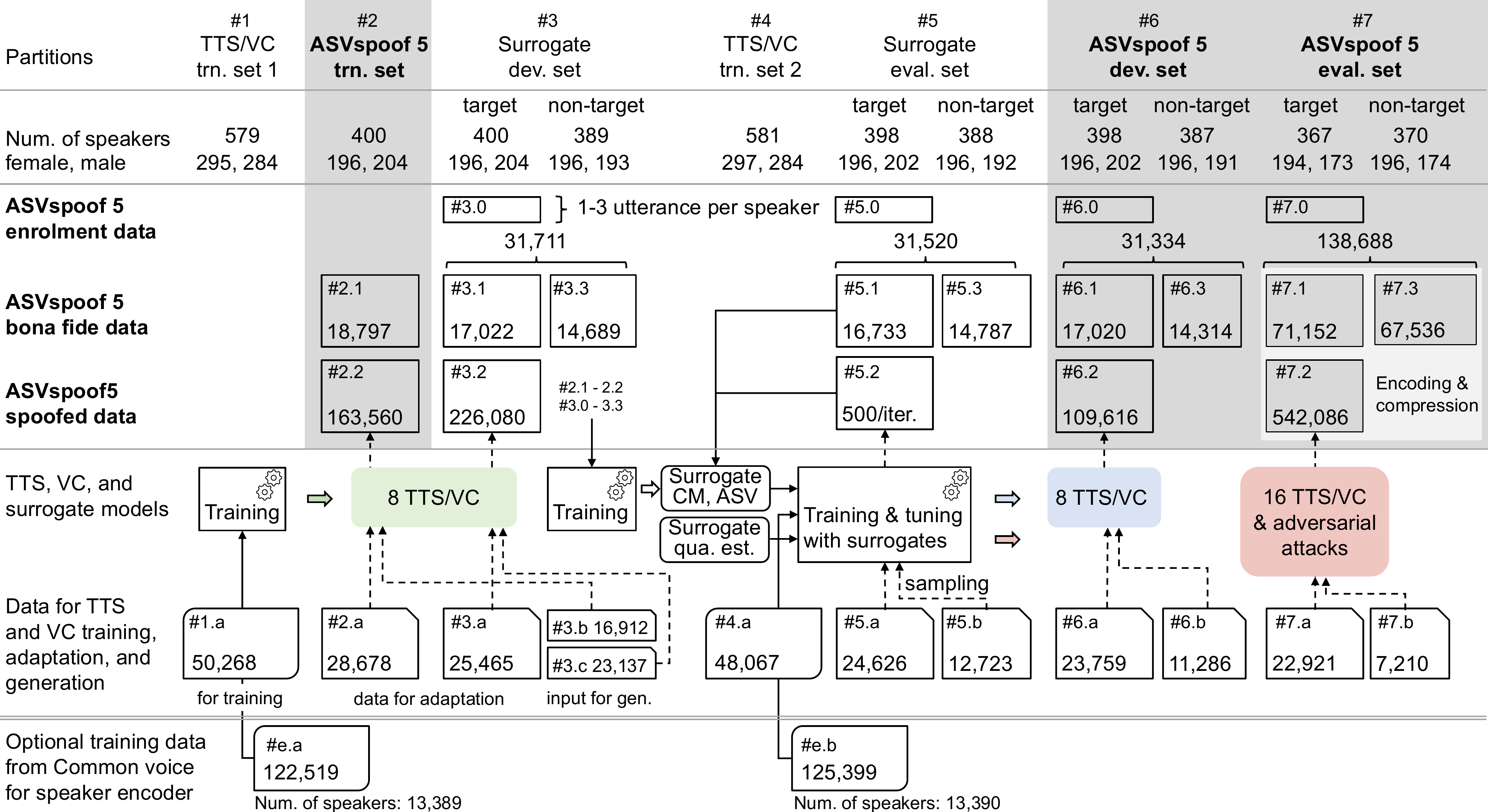}
     \caption{An illustration of the MLS English source database partitioning scheme and its use in generation of the ASVspoof~5 database, including training, development and evaluation sets illustrated in grey boxes. Numbers in the second row indicate the number of speakers in each partition. Those in rectangular blocks denote the number of utterances.  
     The green box denotes the 8 TTS/VC systems used for generation of the ASVspoof~5 training set.  
     The blue box denotes 8 TTS/VC systems used for generation of the development set whereas the red box denotes the 16 TTS/VC systems and adversarial attacks used for generation of the evaluation set.
    Solid black and coloured arrows denote use of data for the training of TTS/VC attacks or surrogate models.  Dashed arrows denote the adaptation towards target speaker and attack generation using input text or source speech utterance.
    }
     \label{fig:ASVspoof5_protocol}
\end{figure*}

At a high level, the ASVspoof~5 database is similar in structure to that of earlier editions: it contains the usual three training, development and evaluation partitions illustrated by  shaded, grey boxes in Figure~\ref{fig:ASVspoof5_protocol}.  The training partition (partition \partnum{2} in Figure~\ref{fig:ASVspoof5_protocol}) is intended for the training of CM and ASV systems as well as spoofing-robust ASV (SASV) systems.\footnote{An SASV system functions in the same way as combined CM/ASV subsystems---they should operate to accept only input utterances which are bona fide and which contain the target speaker voice, and to reject anything else (spoofs/deepfakes and bona fide utterances which do not match the target speaker voice.} It comprises a set of utterances collected from 400 speakers (second row), 19k bona fide utterances (\partnum{2.1}, sourced directly from the MLS database) and 164k spoofed utterances (\partnum{2.2}), generated using a set of 8 distinct TTS/VC algorithms (described in Section~\ref{sec:attacks:trn}).

The ASVspoof~5 development set (\partnum{6}), is intended for the usual purpose and contains two subsets.  The first comprises utterances corresponding to 398 speakers designated as target speakers.  There are between 1 and 3 utterances per target speaker for ASV enrolment (\partnum{6.0}), 17k bona fide utterances (\partnum{6.1}) and 110k spoofed utterances (\partnum{6.2}).  The latter are again generated using a set of distinct TTS/VC algorithms (described in Section~\ref{sec:attacks:dev}) which are different to those used for generation of the training set. The second subset contains 14k bona fide utterances collected from 387 speakers designated as non-target speakers~(\partnum{6.3}).  

Last, the ASVspoof~5 evaluation set, again intended for the usual purpose, has an identical structure to the development set.  The number of speakers in each subset is 367 target speakers and 370 non-target speakers. There are again between 1 and 3 utterances for ASV enrolment (\partnum{7.0}) but a far greater number of bona fide and spoofed utterances, counting 71k and 68k bona fide utterances for target speakers (\partnum{7.1}) and non-target speakers (\partnum{7.3}) respectively.  The 542k spoofed utterances (\partnum{7.2}) are generated using 16 distinct TTS/VC and adversarial attacks (described in Section~\ref{sec:attacks:eval}).  Once again, they are different to those used for generation of the training and development sets.  To simulate a broad variety of applications where speech data is transmitted or compressed, various lossy encoding schemes are also applied to all evaluation data (both bona fide and spoofed). The encoding schemes are described in Section~\ref{sec:post:codec}.

\subsection{Attack generation}
\label{sec:database:attack}
 
In the following we delve deeper into use of the database and protocols for attack generation.  We describe the disjoint MLS database partitions used for generation of spoofed data in the ASVspoof~5 training, development and evaluation partitions. 
Attack algorithms were designed through two phases and in collaboration with external contributors, all experts in TTS/VC technology.%

As has been the case for earlier databases, the ASVspoof~5 database and protocols were designed to support use of the very latest TTS/VC technologies.  Whereas earlier databases were generated using TTS/VC systems which required extensive training data for each target speaker, today's state-of-the-art systems require much less.  Typically, they are trained using huge collections of speech collected from a large number of speakers.  Pre-trained systems are then fine-tuned using relatively few utterances collected from the specific target speaker~\cite{wu15b_interspeech,arik2018neural,yan2021adaspeech,chen2021adaspeech}.  Some systems even operate without fine-tuning and, instead, generate speech in the voice of a specific target speaker by using a single speaker embedding~\cite{jia2018transfer,arik2018neural,cooperZeroshottts2020,wu2022adaspeech4}, i.e.\ zero-shot voice cloning. The ASVspoof~5 database and protocols are hence devised with such TTS/VC systems in mind.

The TTS/VC training set 1 (\partnum{1.a}) is reserved for the training of 8~TTS/VC systems used for the generation of spoofed utterances in the ASVspoof~5 training set (\partnum{2.2}). It contains 50k utterances collected from 579 speakers. Spoofed utterances are generated using held-out subsets of adaptation utterances collected from target speakers and input utterances collected from non-target speakers.  Adaptation is typically used to clone the voices of the 400 target speakers in the ASVspoof~5 training set, for example by fine-tuning towards the voice of a specific target, or by extracting and then using target speaker embeddings for zero-shot voice cloning.  
Input utterances\footnote{For reasons described shortly, input utterances are selected from the disjoint surrogate development partition described in Section~\ref{subsec:surrogates}.} are used for VC or for TTS using corresponding transcriptions.  The choice depends on the specific attack algorithm (described in Section~\ref{sec:attacks}). 

To observe the influence of different durations of adaptation data,
spoofed utterances are generated using different numbers of adaptation utterances.
The configurations are listed in Table~\ref{tab:attackcondition}.\footnote{
Following~\cite{tan2021survey}, we experimented with many different numbers of utterances ranging from 10 seconds to 20 minutes of adaptation data. To maintain a reasonable database size, we retained the reported three configurations which result in substantially different surrogate ASV performance derived using the model described in Section~\ref{subsec:surrogates}.}
The first possible, but technically-demanding scenario involves the use of recordings collected surreptitiously from the target speaker during their interaction with the ASV system (\texttt{AC1}).\footnote{
Utterances segmented from the same audiobook are assumed to be collected in the same recording session.  Audiobook contributors are requested to make recordings without interruptions (see \href{https://wiki.librivox.org/index.php?title=Newbie_Guide_to_Recording}{Librivox recording guidelines}). The session ID is retrieved from the MLS utterance file name.}  These recordings are assumed to be in the order of 10 seconds duration. Their collection in acoustic conditions identical to those for  collection of enrolment data is expected to result in a stronger attack and hence to represent a worst case scenario.
The two additional scenarios reflect the case where the adversary uses short recordings found online ($\approx$30~seconds, \texttt{AC2}) or a large set of utterances from multiple sources collected from a more intensive search ($\approx$20~minutes, \texttt{AC3}).  With these two cases we aim to study the influence of session mismatch,\footnote{Utterances segmented from different audiobooks.} which might result in a weaker attack with respect to \texttt{AC1} as well as the quantity of adaptation data.

\begin{table}[t!]
    \centering
    \caption{TTS/VC adaptation configurations}
    \begin{tabular}{rlll}
        \toprule
         Configuration ID & \texttt{AC1} &  \texttt{AC2} & \texttt{AC3}  \\  \midrule
        Number of adaptation utterances & $\leq$3 & 3 & $\approx$100 \\
        Total duration of adaptation utterances & $\leq$30s & $\approx$30s & $\approx$20m \\ 
        Adaptation and enrolment data from the same session? & $\checkmark$ & $\times$ & $\times$ \\
        \bottomrule
    \end{tabular}
    \vspace{-3mm}
    \label{tab:attackcondition}
\end{table}

The TTS/VC training set 2 (\partnum{4.a}) contains 48k utterances and is used for training of the 16~TTS/VC systems and adversarial attacks used for the generation of spoofed utterances contained within the ASVspoof~5 evaluation set.  
The same set is used for training of the eight~TTS/VC systems for the ASVspoof~5 development set. 
Held-out adaptation and input utterances are again used in similar fashion as for the TTS/VC training set 1.\footnote{Since the ASVspoof~5 development and evaluation sets necessarily include both target and non-target data (to support ASV evaluation), input data for TTS/VC generation can be selected from non-target utterances within the same partition. This is why source utterances are selected from a different partition in case of the ASVspoof~5 train set for which there are no non-target speakers; it is not designed to support ASV evaluation.}

\subsection{Surrogate ASV, CM, and perceptual quality estimation systems}
\label{subsec:surrogates}

Most TTS/VC systems are developed for applications which demand high perceptual quality, whereas adversaries in the context of spoofs/deepfakes might tune TTS/VC systems to increase the likelihood of fooling CM and ASV systems.
The threat of TTS/VC systems used for generation of spoofed utterances contained within the ASVspoof~5 development and evaluation sets is gauged using surrogate CM and ASV systems as well as a perceptual speech quality estimation system. We assume that attackers cannot access the CM and ASV systems used by the defender but can instead choose other open-source implementations as substitutes~\cite{papernot2016transferability}, hence the term `surrogate'.

The surrogate systems are illustrated towards the lower middle of Figure~\ref{fig:ASVspoof5_protocol} and are trained, developed and evaluated in the usual way.  While the provision of surrogate systems\footnote{To reduce the burden on data contributors, access to surrogate systems was provided through an online platform. Contributors could submit generated TTS/VC utterances and use estimated ASV/CM/perceptual quality scores for fine-tuning.} offers an opportunity for attack algorithms to be fine-tuned or optimised, given that spoofed utterances are generated by independent contributors, it is not possible to estimate reliably the relative degree to which each attack is tuned, though some analysis is provided in Section~\ref{sec:analysis:surrogate}.
Invariably, tuning involves merely experimentation to verify the degree to which attacks are effective and to adapt the algorithms accordingly. Surrogate systems were used for fine-tuning attack algorithms that are later used in generation of ASVspoof~5 development and evaluation sets only (not for the train set).

Surrogate CM/ASV systems are trained using ASVspoof~5 bona fide and spoofed utterances (\partnum{2}). The enrolment, bona fide, and spoofed data from partition \partnum{3} are used as surrogate development data.  The structure of the surrogate development (\partnum{3}) and evaluation sets (\partnum{5}), additional disjoint partitions illustrated in Figure~\ref{fig:ASVspoof5_protocol}, is similar to that of the ASVspoof~5 development and evaluation sets.  They are also partitioned into target and non-target subsets.  Each subset contains data collected from approximately 400 speakers.  The surrogate development set contains 17k bona fide utterances (\partnum{3.1}) and 226k spoofed utterances (\partnum{3.2}) for target speakers and 15k bona fide utterances for non-target speakers (\partnum{3.3}).  Spoofed utterances are generated using the same 8 TTS/VC algorithms used for generation of the ASVspoof~5 training set, again using a set of adaptation utterances (\partnum{3.a}) and input utterances for voice conversion, or transcriptions for text-to-speech synthesis (\partnum{3.c}).  

The surrogate evaluation set contains 17k bona fide utterances (\partnum{5.1}) for target speakers and 15k bona fide utterances for non-target speakers (\partnum{5.3}).  TTS and VC systems trained using the TTS/VC training set~2 (\partnum{4.a}) are optionally fine-tuned to increase surrogate CM/ASV system error rates via multiple iterations. In each iteration, TTS/VC systems are used to generate up to 500 spoofed utterances (\partnum{5.2}) after the sampling of adaptation utterances (from \partnum{5.a}) or input utterances for generation (from \partnum{5.b}). The generated spoofed utterances, together with the target and non-target utterances in the surrogate evaluation set (\partnum{5.1} and \partnum{5.3}), are fed to the surrogate models to compute detection error rates which can then be used for TTS/VC system fine-tuning.

\subsection{Speaker and utterance selection}
\label{subsec:spkutt}

The number of utterances per speaker in the source MLS English database is unbalanced (between 1 and 300k utterances per speaker). Such imbalance can lead to bias in the training and evaluation of any derived system.  The partitions illustrated in Figure~\ref{fig:ASVspoof5_protocol} are hence created in deterministic fashion by ensuring that all speakers designated as targets (ASVspoof~5 development and evaluation sets as well as both surrogate sets) have a minimum of two recording sessions. Furthermore, ASVspoof~5 enrolment and bona fide utterances for each target speaker are selected from different recording sessions.  Last, so as to maintain a suitable balance between the influence of speakers for which data is abundant and those for which data is sparse, the number of bona fide utterances for each speaker (target and non-target) is capped at 50 and floored at 30. 

Data for all speakers in the evaluation set, including both targets and non-targets, are removed if data corresponding to the same speakers also appears in the Librispeech database~\cite{panayotov2015librispeech}. Both Librispeech and MLS databases are sourced from LibriVox. Many popular self-supervised learning (SSL) models~\cite[Table IV]{mohamedSelfSupervised2022} are trained using the Librispeech database and use of the same data within the ASVspoof~5 evaluation set would lead to biases in the evaluation of derived systems.

\subsection{Speaker encoder training}

To support the use of  TTS/VC systems that use speaker representations in the form of ASV embeddings to generate outputs in the voice of specific target speakers~\cite{jia2018transfer,cooperZeroshottts2020}, contributors could select to use a subset of the Common Voice (ver. 11.0)~\cite{ardila-etal-2020-common} database for ASV model training.  Similar to the MLS English database, the distribution in terms of data quantity per speaker is unbalanced. To maintain a reasonable balance, but still a suitable number of utterances and speakers, the subset corresponds to the set of speakers for which there is between 10 and 240 seconds of speech. This results in a total of 200 hours of data.  Two different subsets were created, each containing data from approximately 13k speakers.  They contain approximately 123k utterances (\partnum{e.a}) and 125k utterances (\partnum{e.b}), as illustrated in the last row of Figure~\ref{fig:ASVspoof5_protocol}.
Their use is optional and is reserved exclusively for the training of TTS/VC system speaker encoders.

\begin{figure}
    \centering
    \includegraphics[width=1.0\linewidth, trim=0 600 0 0, clip]{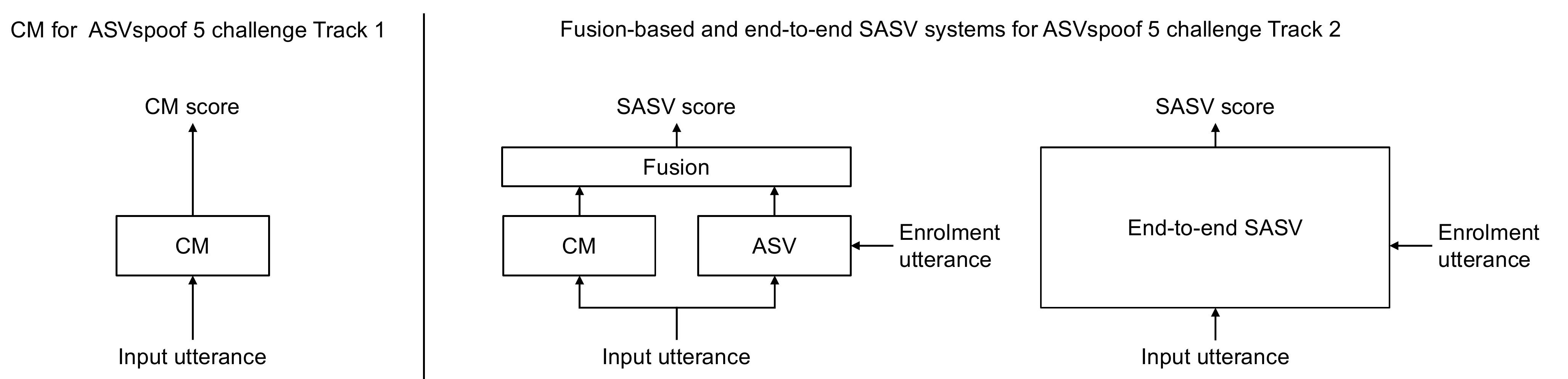}
    \caption{An illustration of the Track~1 spoof/deepfake detection task which involves the design of a CM and the Track~2 spoofing robust automatic speaker verification (SASV) task for which there are two general approaches.}
    \label{fig:cm_sasv}
\end{figure}

\subsection{Challenge tracks}
\label{sec:database:challenge}
The ASVspoof~5 database was designed to support two different tasks which form a pair of challenge Tracks. As illustrated to the left of Figure~\ref{fig:cm_sasv}, Track~1 is a spoof/deepfake detection task and concerns the design of CM systems which produce a score for each input utterance, wherein higher scores indicate a higher likelihood that the input utterance is bona fide (or, equivalently, a lower likelihood that it is spoofed). Track~1 does not involve the design of an ASV system. CMs are implemented using the ASVspoof~5 training (\partnum{2.1}-\partnum{2.2}) and development (\partnum{6.1}-\partnum{6.3}) sets with evaluation then being performed using the evaluation set (\partnum{7.1}-\partnum{7.3}). 

As illustrated to the right of Figure~\ref{fig:cm_sasv},  Track 2 involves the design of SASV systems which produce a score for each input utterance, but now where a higher score indicates a higher chance that the input utterance is bona fide and that it matches the voice in the enrolment utterance, i.e.\ the target speaker (or, equivalently, a lower chance of being anything else).  SASV systems typically take one of two different forms involving either the fusion of independent CM and ASV systems or a single, end-to-end system. SASV systems are implemented and evaluated using the same training, development, and evaluation sets as for Track~1, except that utterances in \partnum{6.0} and \partnum{7.0} are used in addition.  They provide data for the enrolment of target speakers for corresponding trials in the development and evaluation sets. 
Readers are encouraged to consult the ASVspoof~5 challenge evaluation plan~\cite{ASVspoof5_evalplan_phase2} for full details.

\section{Spoofing attacks}
\label{sec:attacks}

\begin{figure}[t!]
    \centering
     \includegraphics[width=\textwidth, trim=0 420 0 0]{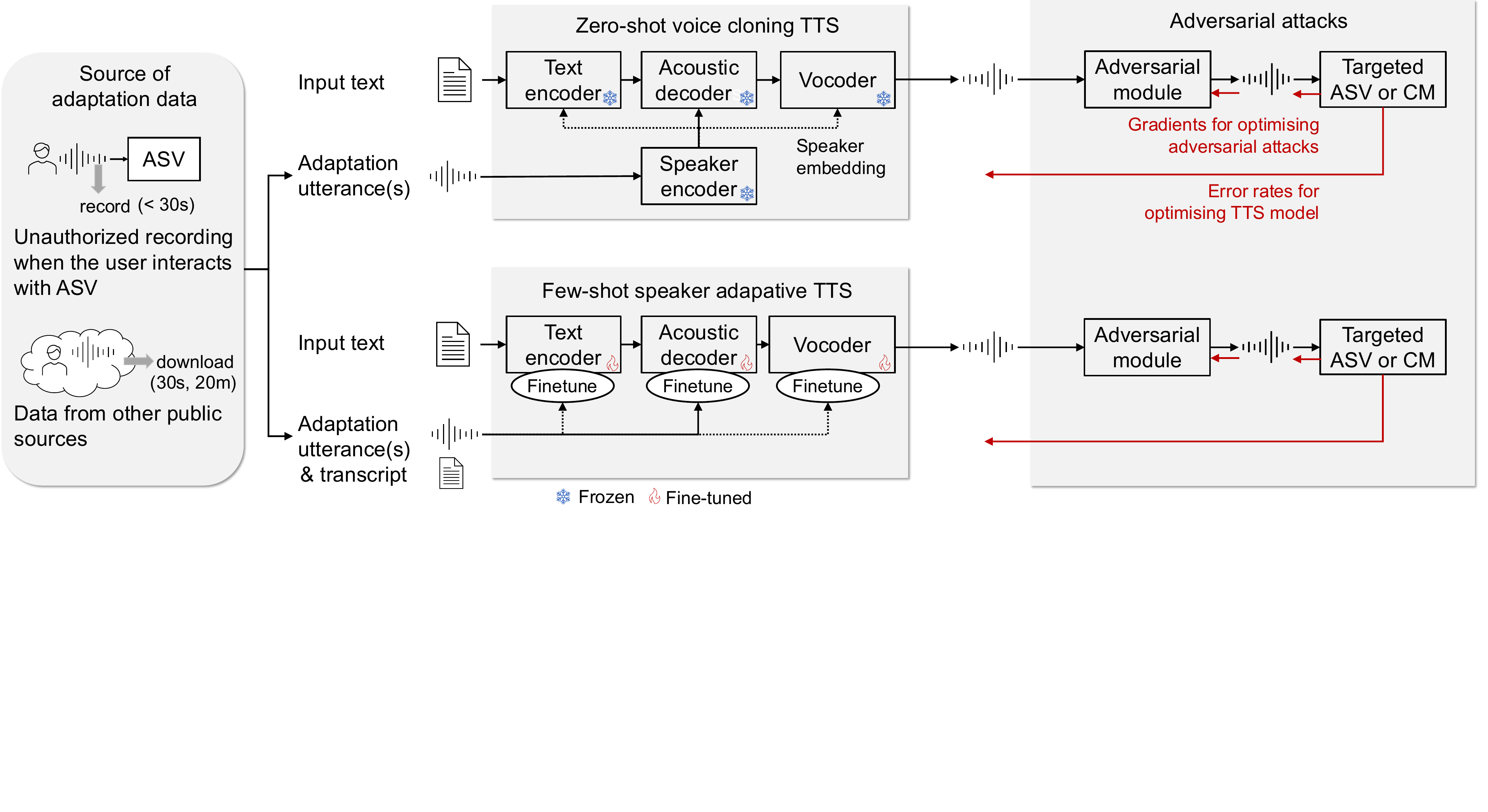}
     \caption{Illustration of speech generation using zero-shot (top-middle) and few-shot (bottom-middle) TTS systems. Few-shot TTS modules are fine-tuned using the adaptation data before generation, while zero-shot TTS modules are frozen (not fine-tuned). Dotted lines indicate the optional use of fine-tuning or speaker embeddings by the text encoder and vocoder. Illustrated to the left is the source of adaptation utterances for a target speaker. 
     Illustrated to the right is the creation of adversarial attacks, 
     whereby a TTS system is optimised using either the gradients or the error rates returned by the targeted ASV or CM system. 
     The red lines indicate operations applied only for training. 
     A similar figure can be plotted for VC attacks by replacing the input text, the text encoder, and the acoustic decoder with an input utterance from a non-target speaker, a waveform encoder, and a VC-oriented acoustic decoder, respectively.}
     \label{fig:tts_overview}
\end{figure}

The TTS/VC research landscape has evolved dramatically since the release of the most recent, previous ASVspoof database. New techniques borrowed from the deep learning community, e.g., diffusion, have been used to create even higher quality synthetic speech. However, successful spoofing attacks are not necessarily implemented using the latest techniques, and might not even require the highest quality synthetic speech to fool a CM~\cite{warren2024better} or an ASV system~\cite{jung_what_2024}. It is hence imperative that attacks generated with both recent and legacy TTS/VC algorithms are included in the ASVspoof 5 database. 
Also included are the latest adversarial attacks.
While it is impractical to cover all related TTS/VC architectures and adversarial attacks, we considered those to the extent possible %
and collected a database of attacks generated with a set of 32 attack algorithms listed in Table~\ref{tab:attacks}. 

Before describing each attack, we first provide a general definition of TTS, VC and adversarial attacks. Readers who are already familiar with ASVspoof databases and associated generative technologies can choose to skip this material and move directly to Section~\ref{sec:attacks:trn} which provides a description of the attacks used in creating the training, development, and evaluation sets.

\subsection{Attack types}
\textbf{TTS-based attacks} are performed using a synthetic utterance which contains the voice of a target speaker learned from one or more adaptation utterances and the linguistic content of a provided text input.  As illustrated in Figure~\ref{fig:tts_overview}, TTS systems follow one of two general approaches, namely zero-shot voice cloning or few-shot speaker adaptive TTS.\footnote{
Some legacy approaches such as unit-selection TTS are neither zero-shot nor few-shot in strict terms. 
To help clarify the concept, we consider the operation of the unit-selection TTS, which is used in the ASVspoof~5 database, as a special form of few-shot speaker-adaptive TTS. 
These systems synthesize speech by selecting and concatenating small speech segments called units. The acoustic decoders (i.e., the selection algorithms) extract units from the adaptation utterance(s). The resulting sequence of units is then simply concatenated by the `vocoder'. Readers are encouraged to refer to TTS literature for an in-depth explanation of legacy approaches~\cite{taylor2009text}.} 
A text encoder is used to transform the input text into features which (explicitly or implicitly) encode word pronunciation information and other suprasegmental attributes (e.g., pitch and rhythm). The acoustic decoder then produces acoustic features (e.g., mel-spectrograms), which are further transformed by the vocoder into a speech waveform.\footnote{The modules are not necessarily implemented in the form of DNNs. They can instead be a set of linguistic rules or decision trees. So-called end-to-end TTS systems in the form of a single DNN can be divided conceptually into distinct blocks, each of which corresponds to one of the modules illustrated in Figure~\ref{fig:tts_overview}.} 
The two approaches illustrated in Figure~\ref{fig:tts_overview} differ in how the target speaker voice is learned. In the case of zero-shot voice cloning (top-middle), a target speaker embedding is extracted from the adaptation utterances by a speaker encoder.  
The acoustic decoder, and optionally also the text encoder and vocoder, are then conditioned upon the target speaker embedding.
Since they use target speaker embeddings, zero-shot voice cloning TTS systems are pre-trained using data collected from a large number of speakers, but are not fine-tuned at the level of the target speaker.  This sets zero-shot voice cloning systems apart from few-shot speaker adaptive TTS systems (bottom-middle) which are fine-tuned to the voice of the target speaker using the set of adaptation utterances and their transcriptions.

\textbf{VC-based attacks} are implemented using an input utterance containing the voice of a non-target speaker in place of input text, but again using one or more adaptation utterances which contain the voice of a target speaker.  The voice in the input utterance is converted to that of the target speaker, whereas the linguistic content of the former is preserved.
Approaches to VC can be categorised similarly to the TTS systems illustrated in Figure~\ref{fig:tts_overview}, though with input utterances in place of input text and with a
waveform encoder in place of the text encoder. VC acoustic decoders typically use similar statistical models or deep neural networks (DNNs) as TTS, but with necessary modifications to accommodate input features of different dimension.

\textbf{Adversarial attacks} involve the introduction to an utterance of discreet perturbations designed specifically to increase CM/ASV error rates.  Adversarial attacks usually take the form of subtle, additive noise, e.g., estimated via a fast gradient sign method~\cite{goodfellow2014explaining}, designed for a specific input utterance and targeted CM/ASV system~\cite{liuAdversarial2019,wang2024advsv,li2020adversarial}. These techniques can be costly in terms of computation, especially when adversarial noise is estimated for a large number of utterances. 
Though still designed for a specific CM/ASV system, less costly, more universal adversarial attacks, have also been reported~\cite{panariello23b_interspeech,malacopula}.  While these can be applied to any input, they are applied to already-spoofed utterances with the aim of making them more difficult to detect and are hence designed for the specific, underlying spoofing attack and/or speaker. 
Our use of such adversarial attacks is illustrated to the right of Figure~\ref{fig:tts_overview}, wherein post-processing adversarial modules are applied to spoofed utterances produced by a given TTS/VC system.  Adversarial modules are trainable, convolutive filters which target a specific CM system~\cite{panariello23b_interspeech} or a specific ASV system~\cite{malacopula}. The trainable parameters (i.e., filter coefficients) are learned from the gradients back-propagated from the targeted CM/ASV system. %
The fine-tuning of a TTS/VC system according to the surrogate CM/ASV or perceptual quality estimator outputs (Section~\ref{subsec:surrogates}) is also a form of black-box adversarial training. In this case, the TTS/VC system observes scores or decisions rather than gradients.

\begin{table*}[t!]
    \centering
    \caption{Summary of spoofing attacks in the ASVspoof~5 database. The column Reference lists the name of the TTS system used in other literature, if available. Note that A30 applies both Malafide and Malacopula to the base attack A17. TF: Transformer; CF: Conformer; GST: Globle style token.}
    \small
    \resizebox{\textwidth}{!}{
    \setlength{\tabcolsep}{3pt}{
            \begin{tabular}{lllllllll}
        \toprule
         & & & Category & \shortstack{Text / wav. \\ encoder} & \shortstack{Acoustic \\ decoder} & \shortstack{Speaker \\ encoder} & \shortstack{Vocoder} & Reference \\
        \midrule
        \multirow{8}{*}{\rotatebox{90}{Training set}} & \multirow{8}{*}{\rotatebox{90}{TTS}} 
        & A01 & Zero-shot TTS & TF encoder & Glow & ECAPA& HiFi-GAN & Glow-TTS~\cite{kim2020glow}\\
        && A02 & Zero-shot TTS & TF encoder & Glow & x-vec. & HiFi-GAN  & Glow-TTS \\
        && A03 & Zero-shot TTS & TF encoder & Glow & y-vec. & HiFi-GAN  & Glow-TTS \\
        && A04 & Zero-shot TTS & TF encoder & Diffusion & ECAPA & HiFi-GAN  & Grad-TTS~\cite{popov2021grad} \\
        && A05 & Zero-shot TTS & TF encoder & Diffusion & x-vec. & HiFi-GAN   & Grad-TTS \\
        && A06 & Zero-shot TTS & TF encoder & Diffusion & y-vec. & HiFi-GAN   & Grad-TTS \\
        && A07 & Zero-shot TTS & TF encoder & TF encoder & ECAPA & HiFi-GAN & FastPitch~\cite{l2021fastpitch} \\
        && A08 & Zero-shot TTS & TF encoder & Normalizing flow & x-vector & HiFi-GAN & VITS~\cite{kim2021vits}\\
        \midrule
        \multirow{8}{*}{\rotatebox{90}{Development set}} & \multirow{8}{*}{\rotatebox{90}{TTS and VC}} 
        & A09 & Zero-shot TTS & CF & CF + Glow, F0 & GST & HiFi-GAN & ToucanTTS~\cite{lux2023toucantts}\\
        && A10 & Zero-shot TTS & CF & CF + Glow & GST & HiFi-GAN v2 & ToucanTTS \\
        && A11 & Zero-shot TTS & CNN + RNN & Attention + AR  & RNN-G2E & WaveGrad & Tacotron2~\cite{shen2018natural} \\
        && A12 & Few-shot TTS & Linguistic-based & Unit (phone) select & - & Wav. concat. & - \\
        && A13 & Zero-shot VC & CNN & CNN & GRU-RNN & WaveGlow & StarGAN-ZSVC~\cite{mathew2021} \\
        && A14 & Zero-shot TTS & TF encoder & Normalizing flow & RawNet3 & HiFi-GAN & YourTTS~\cite{casanova2022yourtts} \\
        && A15 & Few-shot VC &  CNN-VAE & CycleGAN & - & WaveNet &  \cite{albadawy20_interspeech} \\
        && A16 & Zero-shot VC & wav2vec + F0 & - & CAM++ & HiFi-GAN & - \\
        \midrule
        \multirow{17}{*}{\rotatebox{90}{Evaluation set}} & \multirow{9}{*}{\rotatebox{90}{TTS and VC}}
        & A17 & Zero-shot TTS & BERT + TF encoder & VQ-VAE & ECAPA & HiFi-GAN & ZMM-TTS~\cite{gong2023zmm} \\
        && A19 & Few-shot TTS & Linguistic-based & Non-uniform unit-select & - & Wav. concat. & MaryTTS~\cite{schroder11interspeech} \\
        && A21 & Zero-shot TTS & CF & CF + Glow, F0 & GST & BigVGAN & ToucanTTS \\
        && A22 & Zero-shot TTS & CF & CF + Glow & GST & BigVGAN & ToucanTTS \\
        && A24 & Zero-shot VC & RNN-CNN PPG & CNN & ECAPA & HiFi-GAN & - \\
        && A25 & Zero-shot VC & TF encoder & Duffsion & RNN-G2E & HiFi-GAN & DiffVC~\cite{popov2021diffusion} \\
        && A26 & Zero-shot VC & wav2vec + F0 & - & CAM++ & HiFi-GAN & - \\
        && A28 & Zero-shot TTS & TF encoder & Normalizing flow & H/ASP & HiFi-GAN & YourTTS, pre-trained \\
        && A29 & Zero-shot TTS & GPT-2 &  - & H/ASP & HiFi-GAN & XTTS~\cite{casanova2024xtts}, pre-trained \\
        \cmidrule{2-9}
        & \multirow{8}{*}{\rotatebox{90}{Adversarial}} & & Category & Base attack & Targeted model & Filter length & \#. branches & \\ 
        \cmidrule{3-9}
        && A18 & Malafide & A17 (TTS) & CM (AASIST) & $L=1025$ & - \\
        && A20 & Malafide & A12 (TTS) & CM (AASIST) & $L=1025$ & - \\
        && A23 & Malafide & A09 (TTS) & CM (AASIST) & $L=1025$ & - \\
        && A27 & Malacopula& A26 (VC) & ASV (CAM++) & $L=1025$ & $K=5$\\
        && A30 & Malafide & A17 (TTS) & ASV (CAM++) & $L=1025$ & -\\
        &&     & +Malacopula &  &  & $L=\phantom{0}257$ & $K=3$\\
        && A31 & Malacopula & A22 (TTS) & ASV (CAM++) & $L=\phantom{0}513$ & $K=3$\\
        && A32 & Malacopula & A25 (VC) & ASV (CAM++) & $L=1025$ & $K=5$\\
        \bottomrule
        \end{tabular}
    }
    }\vspace{-3mm}
    \label{tab:attacks}
\end{table*}

\subsection{Training set}
\label{sec:attacks:trn}
The set of attack algorithms used to generate spoofed data contained within the ASVspoof~5 training set are all forms of zero-shot voice cloning TTS (top-middle of Figure~\ref{fig:tts_overview}). 
A summary of each is presented in the top-most block of Table~\ref{tab:attacks}. More detailed descriptions are presented in the following.

\textbf{A01}: a neural TTS system based on Glow-TTS~\cite{kim2020glow}. At the core is an acoustic decoder which generates mel-spectrograms via a normalising-flow-based deep generative model named Glow~\cite{kingma2018glow}. The acoustic decoder is conditioned on latent linguistic features which are extracted from the input text by a text encoder using the Transformer encoder architecture~\cite{vaswani2017attention}. The vocoder is a generative adversarial network (GAN)-based DNN named HiFi-GAN~\cite{kong2020hifigan}.  
The speaker encoder is an ECAPA-TDNN~\cite{desplanques2020ecapa}. 
Glow-TTS is open-sourced and is widely used within the TTS community. 

\textbf{A02}: identical to A01, except for use of a ResNet-34-based speaker encoder~\cite{he2016deep}.

\textbf{A03}: identical to A01, except for use of a TDNN-Y-vector-based speaker encoder~\cite{zhu21b_interspeech}.

\textbf{A04}: a neural TTS system based on Grad-TTS~\cite{popov2021grad}, featuring a diffusion-based acoustic decoder. The acoustic decoder generates a mel-spectrogram via an iterative reverse diffusion process~\cite{song2020score}. 
From latent linguistic features generated by the text encoder, the acoustic decoder with a U-Net architecture~\cite{ronneberger2015u} is used to predict a `residual' (or, more specifically, the gradient field which maximises the probability of data to be generated). The residual is then summed with the latent features, the output of which is again fed to the acoustic decoder to produce another residual. The process is repeated ten times, and the output of the last iteration is used as a generated mel-spectrogram.
The text encoder is a Transformer encoder with text and speaker embedding inputs. The vocoder is a HiFi-GAN. 
The implementation is based on the code released by the original paper, but speaker embeddings are extracted using an ECAPA-TDNN speaker encoder.

\textbf{A05}: identical to A04, except for use of a ResNet-34-based speaker encoder~\cite{he2016deep}.

\textbf{A06}: identical to A04, except for use of a TDNN-Y-vector-based speaker encoder~\cite{zhu21b_interspeech}.

\textbf{A07}: a neural TTS system based on FastPitch~\cite{l2021fastpitch} which uses a feedforward DNN without a recurrent layer to reduce generation time. The text encoder uses a stack of Transformer feedforward blocks to convert the input text into a sequence of latent linguistic feature vectors. The acoustic decoder then uses a convolutional neural network (CNN) to predict pitch for each latent vector. 
The paired pitch predictions are transformed to the same dimension as the latent vector before they are summed together. The summed vectors are then fed to another stack of Transformer feedforward blocks to generate a mel-spectrogram. 
A HiFi-GAN-based vocoder is then used to convert the mel-spectrogram into a speech waveform. 
The speaker encoder is an ECAPA-TDNN.
FastPitch is known for its fast generation speed.

\textbf{A08}: a variational inference with adversarial learning for end-to-end text-to-Speech (VITS)~\cite{kim2021vits} system 
which uses a Transformer-based text encoder and a HiFi-GAN vocoder. The acoustic decoder is based on normalising flow, similar to A01-A03. 
However, the text encoder, the acoustic decoder, and the vocoder are jointly optimised by maximizing a variational evidence lower bound~\cite{kim2021vits}. 
Speaker embeddings are x-vectors~\cite{Snyder2018XVectorsRD}.
VITS produces especially high quality synthetic speech and is readily accessible with the open-sourced toolkit ESPNet~\cite{hayashi2020espnet}. 

\subsection{Development set}
\label{sec:attacks:dev}
The ASVspoof~5 development set contains attacks generated with a mix of zero-shot and few-shot TTS and VC systems. A summary of each is presented in the second block of Table~\ref{tab:attacks}.

\textbf{A09}: a zero-shot TTS system~\cite{lux2023toucantts} implemented using the IMS Toucan speech synthesis toolkit \cite{lux2021toolkit}. The system is similar to A07 in terms of using feedforward DNNs but features a number of more advanced techniques. Following~\cite{wu2022adaspeech4}, the speaker encoder is equipped with Global Style Tokens (GSTs)~\cite{wang2018gst}, a set of 2,000 latent vectors (or tokens) trained to encode speaker-related information. 
For voice cloning, a CNN is first applied to extract an initial speaker embedding from the adaptation utterance(s). A refined speaker embedding is then produced using a query-key-value attention block~\cite{vaswani2017attention}. The raw embedding serves the query and the GSTs are the keys and values. GSTs are optimised jointly with other system components.
Both the text encoder and acoustic decoder use stacks of Conformer blocks~\cite{gulati2020conformer}, and the acoustic decoder is further supplemented with a Glow-based~\cite{kim2020glow} post-processing module~\cite{ren2021portaspeech}. 
The acoustic decoder is also conditioned on pitch and energy estimates extracted from an input utterance to clone the prosodic patterns of a non-target speaker~\cite{lux2022prosodycloning}.
The vocoder is based on HiFi-GAN.
The Toucan toolkit is available as open-source and generation speed is faster than real time. 

\textbf{A10}: identical to \textbf{A09}, except for the prediction of pitch and energy from the input text using CNNs~\cite{l2021fastpitch} and the use of a smaller HiFi-GAN vocoder model~\cite{kong2020hifigan} than the A07 attack upon which A09 is based.

\textbf{A11}: a zero-shot TTS system based on Tacotron~2~\cite{shen2018natural}. 
The speaker encoder is a three-layer recurrent neural network (RNN) with long-short-term-memory (LSTM) units~\cite{gravesSupervised2008}, trained for a speaker diarization task using generalised end-to-end (GE2E) loss~\cite{wan2018generalized}. 
The text encoder
comprises convolution and recurrent layers. The acoustic decoder uses a location-sensitive attention block~\cite{chorowski2015attention} and an autoregressive (AR) DNN~\cite{shen2018natural} to generate a mel-spectrogram  
which is then converted into a 
speech waveform using the diffusion-based WaveGrad~\cite{chen2020wavegrad} vocoder. 
Tacotron~2 has been reported to produce speech of near-to-natural quality for single-speaker, neutral style TTS~\cite{shen2018natural}. 

\textbf{A12}: a simplified unit-selection TTS system. %
A set of speech units (phones) is constructed for each target speaker.  The set is derived from  the segmentation of the adaptation utterance(s) given the phone alignments produced by a pre-trained automatic speech recognition (ASR) system provided with the Toucan toolkit~\cite{lux2021toolkit}.
This is the only data used; there is no use of any ASVspoof~5 training data, nor of any other externally-sourced training data.
During generation, the system selects and concatenates together sequences of units which match the phonemic symbols of the input text. Phonemic symbols not covered by the pool of units are ignored. If multiple candidate units are available, one is selected at random.
Beyond use of the pre-trained ASR model, the attack makes no use of any additional DNN models and is hence both technically and computationally less demanding than some other attacks.

\textbf{A13}: a zero-shot VC (ZSVC) system based on the StarGAN-ZSVC model~\cite{mathew2021}. The waveform encoder extracts a mel-spectrogram from the input utterance.
The StarGAN~\cite{miyato2018cgans} acoustic decoder then 
converts the mel-spectrogram so that it encodes the same linguistic content but the voice of the target speaker. The speaker embedding is extracted using a stack of recurrent layers using gated recurrent units (GRUs)~\cite{cho-2014}.
The vocoder is WaveGlow~\cite{prenger2018waveglow}, a normalising-flow-based waveform generation model. 
StarGANs are known to perform well in voice conversion tasks~
\cite{kaneko2019} and the StarGAN-ZSVC model is available as open-source~\cite{mathew2021}.

\textbf{A14}: a zero-shot neural TTS system built upon YourTTS~\cite{casanova2022yourtts}, which is itself based on VITS~\cite{kim2021vits}.  A14 is a variant of A08. 
Advances include the higher number of Transformer blocks used in the text encoder and use of a RawNet3 speaker encoder~\cite{jung22_interspeech}. 

\textbf{A15}: a few-shot VC system~\cite{albadawy20_interspeech}. %
The waveform encoder, a CNN-based variational auto-encoder (VAE)~\cite{kingmaAutoencodingVariationalBayes2014}, is used to extract latent linguistic features from the mel-spectrogram of the input speech. They are used by a CycleGAN~\cite{zhu2017unpaired} acoustic model to produce a new mel-spectrogram conditioned on the target speaker embedding. In contrast to other zero-shot VC systems, the acoustic model is fine-tuned to the target speaker using adaptation data and a cycle-consistency training loss~\cite{zhu2017unpaired}. The VAE encoder is nonetheless still pre-trained without fine-tuning.
The vocoder is the CNN-based WaveNet model~\cite{oord2016wavenet}.
CycleGAN-based systems were popular among submissions to the 2020 Voice Conversion Challenge~\cite{Yi2020}.

\textbf{A16}: a zero-shot VC system which uses a voice disentanglement technique~\cite{sun16_interspeech}. The input is first decomposed into speaker embeddings, linguistic latent vectors, and F0 features.
The speaker encoder uses a context-aware masking (CAM++) model~\cite{cam++}, a more efficient time-delay DNN variant 
of the ECAPA-TDNN model.
Linguistic latent vectors are derived using a wav2vec 2.0 base model~\cite{Baevski2020} which is pre-trained using unlabelled data within the VoxPopuli dataset~\cite{wang-etal-2021-voxpopuli} and fine-tuned for ASR using the English subset. 
F0 dynamics are extracted using the Praat toolkit~\cite{parselmouth}. 
During conversion, the speaker embedding is replaced with that of the target speaker. F0 features are linearly scaled to match the first and second statistical moments of those extracted from the adaptation utterance(s). The set of three representations is fed to a HiFi-GAN vocoder to generate a waveform, without use of an acoustic decoder. 
Similar disentanglement-based VC systems were among the top performing submissions to the 2020 Voice Conversion Challenge~\cite{Yi2020}.

\subsection{Evaluation set}
\label{sec:attacks:eval}
Spoofed data in the ASVspoof~5 evaluation set are generated using nine TTS/VC systems and seven adversarial attacks. Three of the TTS attacks, namely A17, A28, and A29, are generated using off-the-shelf systems pre-trained using external datasets. They are included to support the evaluation of detection performance when attack models are well trained using huge quantities of non-matching data. All other systems are trained in the same way as those described above and with the protocols described in Section~\ref{sec:database_overview}.
The TTS/VC systems are described first, followed by the adversarial attacks. 
A summary of each is presented in the two lower-most blocks of Table~\ref{tab:attacks}.

\textbf{A17}: a zero-shot TTS system, dubbed ZMM-TTS~\cite{gong2023zmm}, originally designed for multi-speaker and multi-lingual applications. 
The text encoder comprises a BERT-like module~\cite{nguyenXPhoneBERT2023} and a stack of feedforward blocks. %
The acoustic decoder, which uses a hierarchical vector-quantization VAE (VQ-VAE)~\cite{van2017neural} architecture~\cite{guo2023msmc}, and the HiFi-GAN-based vocoder are jointly optimised. 
Training data is sourced from multiple datasets in six languages.\footnote{The training data includes 2.4k utterances collected from 10 target and 2 non-target speakers,  for whom there is also data in the ASVspoof~5 evaluation set but not overlapped with the 2.4k TTS training utterances. The use of public datasets across multiple application domains is now widespread meaning that data overlap is increasingly more difficult to avoid. 
Bearing in mind that this represents less than 2\% of the speakers in the evaluation set, the overlap simulates a worse-case scenario whereby training is performed using data collected from the target speaker.
Nevertheless, experimental validation (Section~\ref{sec:experiment:asv}) shows that A17 is not more effective in attacking the `seen' target speakers than the `unseen'.
}
The speaker encoder is an ECAPA-TDNN pre-trained using the VoxCeleb~2 dataset~\cite{voxceleb2}.

\textbf{A19}: a classical unit-selection TTS system based on the MaryTTS platform~\cite{schroder11interspeech} and the Voice Building Plugin (v5.4)~\cite{steiner2018creating}. %
The adaptation utterance(s) is used to construct a pool of speech units for each target speaker. A19 is a more sophisticated variant of the A12 unit-selection attack. Speech units are of a non-uniform granularity~\cite{sagisakaSpeech1988}, ranging from individual diphones to segments which span consecutive diphones. 
The unit selection algorithm considers not only whether the selected units match the phoneme sequence of the input text but also the distortion introduced through concatenation~\cite{hunt1996unit}. 
MaryTTS was used in the generation of previous ASVspoof databases (i.e., S10, ASVspoof~2015 and A04/A16, ASVspoof~2019). %
MarryTTS attacks are known to pose a threat to ASV reliability. There is also some evidence that they can be difficult to detect~\cite{Wu-ASVspoof2015,asvspoof2019database,jung_what_2024}, even if speech quality might not be as high as that for some of the more recent neural-based TTS approaches.

\textbf{A21}: a variant of the zero-shot TTS system A09. Enhancements include use of a vocoder based on the Big Vocoding GAN (BigVGAN)~\cite{lee2022bigvgan}, a DNN based upon HiFi-GAN. Trainable periodic activation functions~\cite{ziyin2020neural} are used to improve the quality of voiced sounds and anti-aliasing operations inside the GAN generator are used to reduce high-frequency artefacts in the generated waveforms. %

\textbf{A22}: a zero-shot TTS system identical to A10, except for use of the same BigVGAN vocoder used by A21.

\textbf{A24}: a zero-shot VC system. An RNN-CNN waveform encoder is used to extract a phonetic posteriorgram~\cite{kintzley11_interspeech,sun16_interspeech}, a latent representation of mostly speaker-independent phonetic information. The acoustic decoder, also a CNN, combines the posteriorgram with ECAPA-TDNN speaker embeddings to predict target-speaker-dependent latent features. A HiFi-GAN vocoder is then used to produce speech in the voice of the target speaker.
All components are trained in an end-to-end manner.
A24 is similar to A16 in that input speech is disentangled into speaker-dependent and speaker-independent features. A key difference to A16 is use of GAN-based, end-to-end training. %

\textbf{A25}: a zero-shot, diffusion-based VC system~\cite{popov2021diffusion}. 
The Transformer-based waveform encoder extracts latent features  resembling the mel-spectra of an `average voice', i.e., mel-spectra averaged over the set of speakers in the training data. 
Latent features and speaker embeddings extracted from an RNN-GE2E speaker encoder~\cite{wan2018generalized} are then fed to the acoustic decoder and used as priors for a reverse diffusion process~\cite{song2020score}.
The resulting mel-spectrogram 
is then fed to a HiFi-GAN vocoder for waveform generation.

\textbf{A26}: a variant of the A16 zero-shot VC system. A denoiser~\cite{defossez2020real} is applied to 
the training data. During generation, noise contained in the input is extracted and added to converted speech at the output. 
The addition of background noise may lead to more convincing spoofs/deepfakes and/or act to mask conversion artefacts and hence increase the difficulty of detection.

\textbf{A28}: a pre-trained, zero-shot YourTTS~\cite{casanova2022yourtts} system released with the Coqui toolkit~\cite{Eren_Coqui_TTS_2021}. The training data is in English, Brazilian Portuguese and French languages with no overlap with speakers in the ASVspoof~5 evaluation set. 
Different to A14, also based upon YourTTS, is use of an H/ASP-based speaker encoder~\cite{heo2020clova} pre-trained using the VoxCeleb~2 database~\cite{voxceleb2}. 
The publicly available Coqui implementation of YourTTS is a strong performer for zero-shot TTS tasks~\cite{casanova2022yourtts}.

\textbf{A29}: another pre-trained, zero-shot TTS system, named XTTS~\cite{casanova2024xtts}. Designed for multi-lingual TTS and trained using a mix of four datasets containing speech in 16 languages, it uses the GPT-2 text encoder architecture~\cite{radford2019language}, a HiFi-GAN vocoder, and an H/ASP speaker encoder (same as for A28). 
The XTTS system is also publicly available in the Coqui toolkit. 

\textbf{Adversarial attacks}: The remaining attacks %
are all 
adversarial filtering attacks designed to increase the threat of a selection of TTS/VC systems.
As illustrated to the right of Figure~\ref{fig:tts_overview},  learnable parameters are optimised using gradients which are back-propagated from the compromised CM or ASV model. Once trained, adversarial filtering is applied to spoofed utterances in the evaluation set without further modification. %
There are three classes of attacks which are designed to compromise either CM or ASV systems or their combination.
\begin{itemize}
    \item \textbf{A18, A20, A23}: a set of adversarial attacks which target an AASIST CM (described in Section~\ref{sec:baselines}). The coefficients of an adversarial, non-causal and trainable linear-time-invariant (LTI) filter named Malafide~\cite{panariello23b_interspeech}
    are optimised to increase the scores produced by the targeted CM\footnote{By established ASVspoof convention, higher scores indicate support for the bona fide hypothesis.} via a gradient ascent algorithm. 
    As illustrated in Table~\ref{tab:attacks}, A18, A20 and A23 attacks result from the application of Malafide adversarial filtering with $L=1025$ filter coefficients to utterances produced by A17, A12 and A09 TTS systems.
    \item \textbf{A27, A31, A32}: a set of adversarial attacks which target a CAM++ ASV model.\footnote{The ASV model is the CAM++ model from~\cite{cam++}.  It is chosen only for the reasons that it is different to the baseline ASV system described in Section~\ref{sec:baselines} and that its use hence supports evaluation in a black-box setting.} 
    Speaker-specific filters, named Malacopula~\cite{malacopula}, are composed of $K$ parallel filtering branches, one purely linear and the others non-linear via static power polynomial functions $x^{k}, k\in[1, K]$, all with a trainable, non-causal LTI filter. 
     Filter coefficients are optimised to minimise the difference between a pair of embeddings, the first extracted from a spoofed utterance, the second from an enrolment utterance collected from the target speaker. The opimisation acts to maximise ASV scores for the former and hence to provoke a greater rate of false accept decisions.
    As illustrated in Table~\ref{tab:attacks}, A27, A31 and A32 attacks result from the application of Malacopula filtering with either $L=1025$ or 513 filter coefficients and either $K=3$ or 5 branches to utterances produced by A26, A22 and A25 VC or TTS systems.
    \item \textbf{A30}: an adversarial attack which results from the application of Malacopula filtering with $L=257$ components and $K=3$ branches to utterances produced by the A18 attack which itself results from the application of Malafide filtering to utterances produced by the A17 TTS system.  A30 is the only attack which involves the sequential application of both Malafide and Malacopula filtering.
\end{itemize}

Malafide and Malacopula implementations are identical to those described in~\cite{panariello23b_interspeech} and~\cite{malacopula} respectively which provide comprehensive descriptions of each approach.

\section{Post Processing}
\label{sec:postprocess}

A subset of both bona fide and spoofed utterances are post-processed before being added to the ASVspoof~5 database. 
This includes the application of encoding/compression to utterances in the evaluation set and quality control to reduce shortcut artefacts. 

\subsection{Encoding and compression}
\label{sec:post:codec}

To evaluate detection performance under bandwidth, encoding and compression variation, subsets of data within the ASVspoof~5 evaluation set, including both bona fide and spoofed utterances, were encoded or compressed according to one of the evaluation conditions listed in Table~\ref{tab:codecs}. 
Treatment of the evaluation set only helps to keep the database size manageable while also allowing users to choose their own encoding, compression, or any other data augmentation strategies for training and development pipelines. 
With the aim of maintaining a reasonable database size and facilitating the comparison of performance with and without encoding/compression for the same data, we treated a subset of 20\% of the utterances in the evaluation set.  For each utterance $u$ in this subset, we created an exhaustive set of utterance-codec pairs, i.e., $u$-C00, $u$-C01, $\cdots$, $u$-C11.  For each utterance $v$ in the remaining 80\% of data, we create only a single pair $v$-Cxx wherein Cxx is randomly selected from the 11 conditions in Table~\ref{tab:codecs}.

\begin{table}[h!]
    \centering
    \normalsize
    \caption{Summary of codec and compression conditions in evaluation sets.}
    \setlength{\tabcolsep}{4pt}
    \begin{tabular}{clll}
        \toprule
         & Coding format &  Bandwidth  & Bitrate range (kbit/s) \\ 
        \midrule
        C00 & none & 16 kHz & - \\ 
        C01 & opus & 16 kHz & 6.0 - 30.0 \\ 
        C02 & amr & 16 kHz & 6.6 - 23.05 \\ 
        C03 & speex & 16 kHz & 5.75 - 34.20 \\ 
        C04 & Encodec~\cite{de2023encodec} & 16 kHz & 1.5 - 24.0 \\ 
        C05 & mp3 & 16 kHz & 45 - 256 \\ 
        C06 & m4a & 16 kHz & 16 - 128 \\ 
        C07 & mp3+Encodec & 16 kHz & varied  \\ 
        C08 & opus & \phantom{0}8 kHz & 4.0 - 20.0\\
        C09 & amr & \phantom{0}8 kHz & 4.75 - 12.20\\
        C10 & speex & \phantom{0}8 kHz & 3.95 - 24.60 \\
        C11 & varied & \phantom{0}8 kHz & varied \\
        \bottomrule
    \end{tabular}
    \vspace{6mm}
    \label{tab:codecs}
\end{table}

Condition C00 serves as a reference for which there is no encoding or compression.
For conditions C00-C07, all data is sampled at a rate of 16~kHz. In contrast, conditions C08-C11 correspond to a narrow band setting under which all data is down-sampled to 8~kHz.
Conditions C01-C11 all involve the application of lossy encoding/compression algorithms at the bitrates indicated in Table~\ref{tab:codecs}.
C04 and C07 use the recent, deep-learning-based Encodec~\cite{de2023encodec} scheme.
For C07, mp3 encoding is applied prior to Encodec to simulate a channel with double compression. %
Finally, C11 is an experimental setting in which utterances are linearly convolved with the pre-computed, non-linear responses of a set of eight end-to-end calling pipelines, namely calls made from a device to a call center platform over a public switched telephone network (PSTN). 
Responses were estimated using the synchronised swept sine approach described in~\cite{novak2015synchronized} using one of six different calling devices and one of four different audio injection methods. Specific configurations for C11 are detailed in~\ref{app:codec11}.

All data is distributed with a common sampling rate of 16~kHz;
data initially downsampled to 8~kHz in conditions C08-C11 
are upsampled to 16~kHz after encoding/compression. Although upsampling does not increase the effective bandwidth, distribution of data at a common sampling rate reduces the risk that users treat data sampled at 8~kHz with a detector trained using data sampled at 16~kHz.

\subsection{Shortcut artefacts}
Shortcut artefacts are semantically unrelated to the spoofing/deepfake detection problem but can nonetheless be utilised for detection.  Usually the result of dataset collection or generation procedures, the problem of shortcut artefacts can be particularly insidious.  When assessed in the laboratory using evaluation datasets contaminated by such artefacts, detection performance can be strong~\cite{lapuschkin2019unmasking}.  But, when these same systems are deployed in the wild, conditions in which database artefacts cannot be relied upon, performance may degrade catastrophically. 

To illustrate the problem, consider a set of TTS systems which set the peak waveform amplitude to specific, pre-set levels.  The peak amplitude is then an informative indicator of the content having been generated by one of these same TTS systems.  A detector which exploits the peak amplitude shortcut might perform well when it is assessed using similarly generated data, but will likely fail if, for example, a TTS system is configured to generate outputs with random peak amplitude.
In reality, the peak amplitude is semantically unrelated to the detection problem; an utterance generated using TTS synthesis is a spoof/deepfake \emph{regardless} of the peak amplitude. 

\begin{figure*}[!t]
    \centering
    \includegraphics[width=\textwidth]{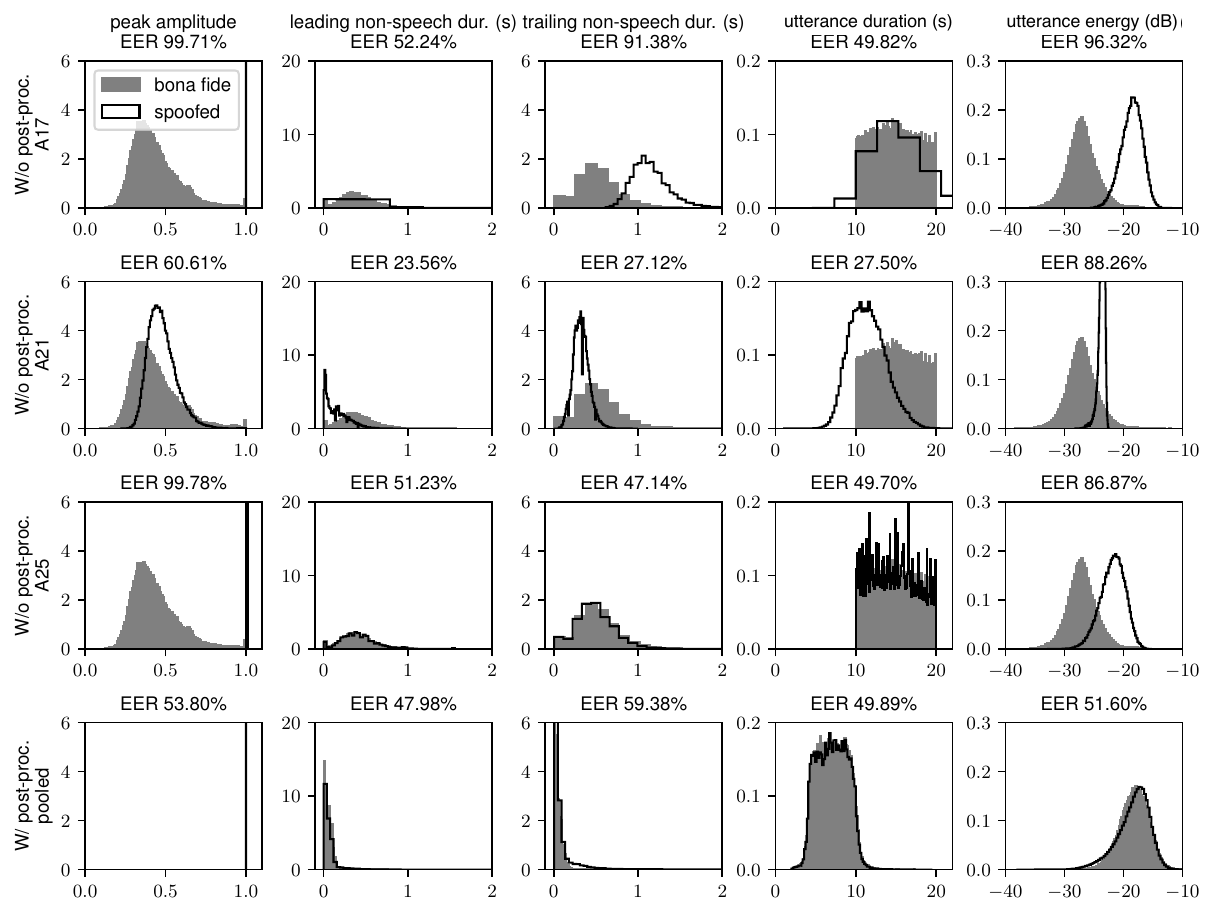}
    \caption{Distributions of potential shortcut artefacts (different columns) \textbf{without (top three rows) and with post-processing (bottom row)}. Spoofed data without post-processing are from A17 (TTS), A21 (TTS), and A25 (VC). The bottom row shows the distribution from all the evaluation data after post-processing. The equal error rate (EER) shown at the top of each sub-figure is computed using the shortcut artefacts of the bona fide and spoofed utterances. An EER closer to 50\% suggests a greater overlap between the distributions of the two classes.}
    \label{fig:statistics_dur}
\end{figure*}

We created a post-processing pipeline to detect and reduce the impact of potential shortcut artefacts.  It was applied to the ASVspoof~5 evaluation and development sets, but \emph{not} to the training set. This choice encourages database users to explore techniques which are able to extract the most semantically relevant cues while also avoiding the pitfalls of shortcut learning. 
Our analysis revealed five potential shortcut artefacts: the peak waveform amplitude; the durations of the leading and trailing non-speech segments~\cite{chettriDataset2020,muller21_asvspoof}; the duration and energy of the whole utterance. The non-speech and speech segments of the utterance in the ASVspoof~5 evaluation set are annotated using the Whisper system~\cite{radford2023robust}.
Figure~\ref{fig:statistics_dur} shows the distribution of each shortcut artefact for bona fide utterances (black profiles) and spoofed/deepfake utterances (grey profiles) before post-processing.  Distributions are shown in the first three rows for utterances generated using A17 (TTS), A21 (TTS), and A25 (VC) attacks. 
They show that, for A17 and A25 attacks, the peak amplitude is rescaled to 1.0, a distinct difference to the distribution for bona fide utterances.  
Whereas there is little difference between the distributions of leading non-speech segment durations for bona fide and spoofed/deepfake utterances for A25, they tend to be of shorter duration for A21.  For A17, the duration of leading non-speech segments is uniformly distributed, even if the range is similar to that of bona fide utterances.
Even so, the range in the duration of trailing non-speech segments for A17 differs to that for bona fide utterances. %
The duration of trailing non-speech segments for A21 is generally again shorter than those for bona fide utterances. %
The distribution of total duration for A25 overlaps with that of bona fide data. This is because A25, a VC system, does not alter the total duration of the input utterance. In the case of TTS-based attacks A17 and A21, the total duration is controlled by the acoustic model, hence the difference between distributions for bona fide and spoofed/deepfake utterances in these cases. There are also distinct differences in distributions of average energy for all three attacks.  %

The post-processing pipeline is designed to reduce the differences between the artefact distributions for spoofed/deepfake and bona fide utterances. %
Let us assume an input waveform $\boldsymbol{x}=(x_1, x_2, \cdots, x_N)$ with $N$ sampling points.
First, the waveform is linearly scaled by a factor $r\in\mathbb{R}^{+}$ so that the peak amplitude $\max_n r\|{x}_n\|$ is equal to 1.0. 
Next, the duration of leading and trailing non-speech segments are trimmed, after annotation of segment boundaries using the Whisper system if the input is from the evaluation set, or an energy-based speech activity detector~\cite[\S 5.1]{kinnunen2010overview} for the development set utterances. 
Given the annotated leading non-speech segment $(x_1, x_2, \cdots, x_{N_s})$ with $N_s$ samples, an index $\tilde{n}_s$ is drawn at random from the probability mass function $\text{Pr}(n)=\frac{\exp(-|x_n|)}{\sum_{i=1}^{N_s}{\exp(-|x_i|)}}, n\in[1, N_s]$. The sub-segment $(x_1, x_2, \cdots, x_{\tilde{n}_s})$ is then trimmed.\footnote{The time resolution of the segment boundary produced by the Whisper system (and the energy-based speech activity detector) is 10~ms on account of the stride of the input mel-spectrogram. The value of $N_s$ is hence a multiple of 160 (= 10 ms $\times$ 16~kHz). We draw $\tilde{n}_s$ from $\text{Pr}(n)$ so that there is randomness in the location of the trimming. $\text{Pr}(n)$ gives a higher chance to select the index at samples where the waveform amplitude is low.}
For the trailing non-speech segment, an index $\tilde{n}_e$ is drawn at random %
in the same way so that the sub-segment $(x_{\tilde{n}_e}, x_2, \cdots, x_{N})$ is trimmed.
Finally,
a float-valued duration is drawn at random from a uniform distribution between 4.0 and 10.0 seconds.
A contiguous chunk of this duration is then selected from within the remaining speech segment as the pipeline output.
There is no further scaling of the utterance energy since 
we observed similar energy distributions at the pipeline output for bona fide and spoofed/deepfake utterances without additional normalisation.
All data is distributed in FLAC format with a sampling rate of 16~kHz. Immaterial metadata, such as the audiobook title, is removed lest it also serve as a shortcut to help distinguish between spoof/deepfake and bona fide utterances.

Plotted in the bottom row of Figure~\ref{fig:statistics_dur} are distributions for each 
of the shortcut artefacts for the full evaluation set after post-processing, again for spoofed/deepfake and bona fide utterances. 
Post-processing reduces the discrepancies substantially. 
Even though the average energy is not manipulated directly, the distributions post-processing are near-to-identical, as they are too for each of the shortcut artefacts. %
Fully overlapping distributions indicate that post-processing successfully reduces shortcut artefacts linked to peak amplitude, non-speech and full utterance durations as well as average energy and provide assurances that they are unlikely to provide cues relevant to detection.

\begin{figure}[t!]
    \centering
    \includegraphics[width=\textwidth, trim=20 0 0 0]{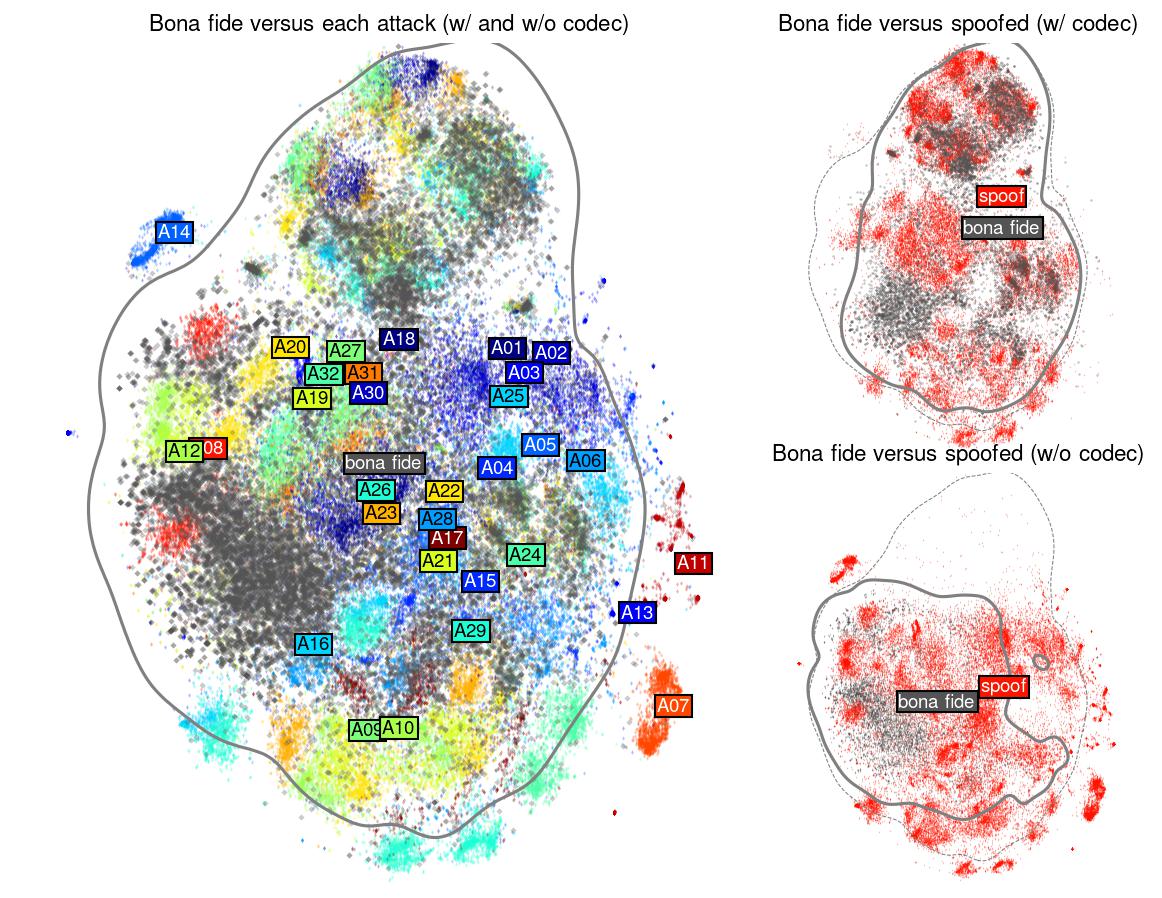}
    \caption{Illustration of bona fide and spoofed samples from each attack in a t-SNE embedding space (sub-figure on the left). Dots in a grey colour correspond to the bona fide samples; others are spoofed samples. The contour encloses 98.9\% of the probability mass of the bona fide samples estimated using a kernel density estimate method. The label of each data class is positioned at the mean of its corresponding data distribution. 
    As alternative views, the top-right and bottom-right sub-figures, which plot all the spoofed samples in the same colour, show the data with and without codec/compression, respectively. The contours in solid lines encloses 98.9\% bona fide samples in the sub-figures. For comparison, the contour from the sub-figure on the left side is plotted as dotted lines in the sub-figures on the right side.
    }
    \label{fig:tsne}
\end{figure}

\section{Visualisation}
\label{sec:visualize}

We present a visualisation of the database characteristics using t-distributed stochastic neighbour embedding (t-SNE) plots~\cite{van2008visualizing} and a hierarchical clustering dendrogram~\cite[Section 14.3]{hastie2009elements}. The objective is to highlight similarities and differences between bona fide and spoof/deepfake utterances.  %

Illustrated in Figure~\ref{fig:tsne} is a representation of the ASVspoof~5 database in the form of a 2-dimensional t-SNE plot~\cite{van2008visualizing}.  Each point corresponds to an  embedding extracted using an ASV system (the baseline system described in \S~\ref{sec:experiment}).  Grey points correspond to bona fide utterances while coloured points correspond to one of the 32 attacks. To reduce computation time and visual clutter, points are shown for a random selection of 10\% of the utterances in the training, development and evaluation sets, with and without encoding/compression for the evaluation set only.  The recipe used to produce Figure~\ref{fig:tsne} is the same as that used to produce similar plots in~\cite{Wu2017-ASVspoof-IEEE-J-STSP} and~\cite{asvspoof2019database} and involves attack-level, within-class covariance normalisation, length normalisation and speaker-level data whitening. 
Also plotted as a grey solid line in Figure~\ref{fig:tsne} is a confidence contour which encloses 98.9\% of the probability mass of the bona fide data distribution. It is estimated using a kernel density estimate (KDE) method.\footnote{We use the \href{https://seaborn.pydata.org/generated/seaborn.kdeplot.html}{ Seaborn KDE API}~\cite{Waskom2021}. The confidence level of 98.9\% was used in deriving a similar visualisation for the ASVspoof 2019 database~\cite[Figure 3]{asvspoof2019database}, though in the form of a confidence ellipse. %
} 

Compared to a similar plot for the ASVspoof 2019 database presented in~\cite[Fig. 3]{asvspoof2019database}, a greater proportion of the attacks lie within the confidence contour. However, attacks A07, A11, and A14 lie outside of the contour, suggesting distinct differences to bona fide utterances. A13, another outlier, collapses into a cluster with low variance, indicating a lack of inter-speaker variation.
Illustrated to the right in Figure~\ref{fig:tsne} is a pair of additional t-SNE plots, the difference between which shows the impact of encoding/compression. %
The contours are plotted in the same way as before.
The overlap between bona fide and spoofed/deepfake utterances is 
greater with encoding/compression than without, indicating a greater challenge to detect spoofed/deepfake utterances.

Another expected finding is the apparent similarity between  spoofing attacks based on the same generative technology, for example, \{A01, A02, A03\}, all neural TTS systems based on Glow-TTS, \{A04, A05, A06\}, all based on Grad-TTS, and \{A09, A10\}, both zero-shot TTS systems implemented using the IMS Toucan speech synthesis toolkit. Attacks \{A01, A02, A03\} (and \{A04, A05, A06\}) share the same acoustic model and vocoder, but use different types of speaker embedding. The use of different approaches to predict prosodic parameters (i.e., F0, energy, and duration) does not appear to cause substantial differences between A09 and A10.  Use of a different vocoder architecture in an otherwise similar attack, as is the case for A09 versus A21 and A10 versus A22, leads to more substantial differences. This observation suggests that, at least in this case, the vocoder has a greater influence on the differences between bona fide and spoofed/deepfake utterances than the approach to predict prosodic parameters. 
More detailed analysis is complicated by the manifold differences between each attack algorithm.

\begin{figure}[t!]
    \centering\includegraphics[width=0.95\textwidth]{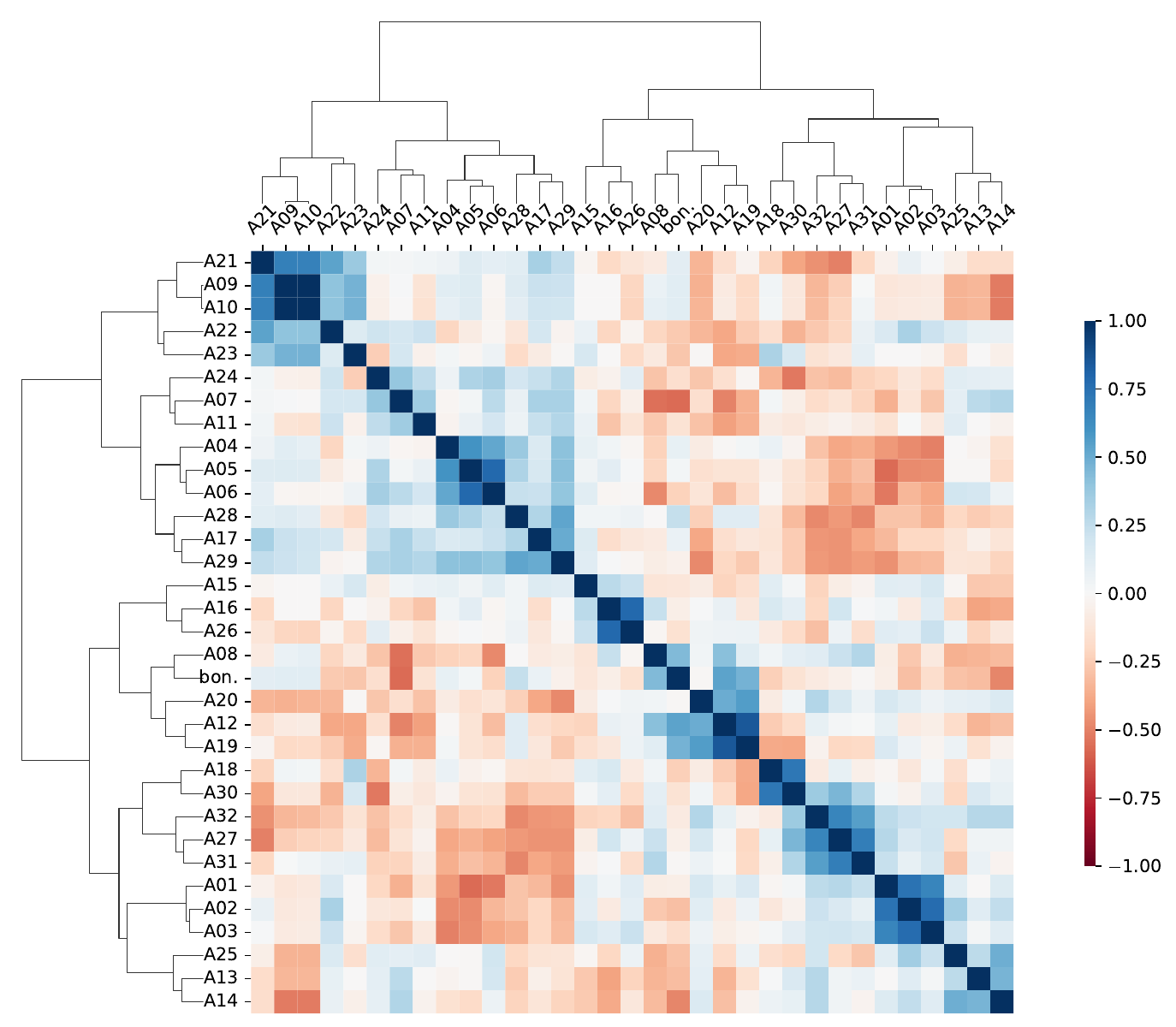}
    \caption{Dendrogram (hierarchical cluster map) of attacks in the ASVspoof 5 database. For visualisation, heatmap plots the cosine similarity between the each pair of the speaker embeddings subsets. Implementations of the dendrogram and cosine similarity are described in Appendix~\ref{app:dendrogram}.}
    \label{fig:embedding-dendrogram}
\end{figure}

Figure~\ref{fig:embedding-dendrogram} provides an alternative visualisation of the database in the form of a \emph{dendrogram} obtained using an agglomerative hierarchical clustering approach, similar to that presented in~\cite[Figure 3]{asvspoof2019database} for the ASVspoof 2019 database. The dendrogram is derived from the same set of speaker embeddings used in generating the t-SNE visualisation. Implementation details are described in~\ref{app:dendrogram}. The clustering is based upon pairwise cosine similarities between embeddings extracted from bona fide utterances and spoofed/deepfake utterances generated using each of the 32 attacks, a visualisation of which is depicted in the colour-scaled heatmap also shown in Figure~\ref{fig:embedding-dendrogram}.  The most similar utterances are indicated in dark blue, whereas the most different are indicated in dark red.

The dendrogram (and the similarity heatmap) show similar clusters of attacks as the t-SNE plot, e.g., \{A01, A02, A03\} and \{A04, A05, A06\}. The high similarities within these two groups of TTS systems may be because the three systems within each group use the same acoustic decoder, text encoder and vocoder, even if they use different speaker embedding extractors. However, the two sub-groups have low inter-group similarity -- they are located in different branches of the dendrogram tree. We observe a similar clustering of \{A09, A10, A21, A22\}, all implemented using the ToucanTTS toolkit.

It is of interest to see whether or not the similarities between bona fide and spoofed/deepfake utterances and between spoofed/deepfake utterances generated with different algorithms correlate with those observed from automatic detection results.  A discussion of this correlation is presented later in Section~\ref{sec:experiment}. We nonetheless stress that the visualisations presented above are derived using \emph{oracle} speaker and attack labels which are unavailable during detection.

\section{TTS and VC system optimisation using surrogate models}
\label{sec:analysis:surrogate}

We describe the optimisation of a subset of TTS and VC systems using the surrogate models introduced in \S~\ref{subsec:surrogates}.  %
Surrogates are based on open-sourced implementations of popular ASV and CM architectures. 
The surrogate ASV system is an ECAPA-TDNN~\cite{desplanques2020ecapa} model with %
cosine scoring implemented using the SpeechBrain toolkit~\cite{ravanelli2024open}. The system is trained using the VoxCeleb1~\cite{nagrani2017voxceleb} training set and VoxCeleb2~\cite{voxceleb2} development set, with data augmentation based upon additive noises from the MUSAN dataset~\cite{musan2015} and reverberation from the RIR dataset~\cite{Ko2017ASO}. 
Surrogate CM systems include AASIST~\cite{jung2022aasist}, RawNet2~\cite{tak2021end}, and LCNNs with LFCC features~\cite{asvspoof2019}, all implemented using source codes provided by the respective authors.
Training is performed from scratch using the ASVspoof~5 training set. %
The best checkpoint is selected based on the lowest EER for the surrogate development set.

\begin{figure}[t!]
    \centering
     \includegraphics[width=\textwidth, trim=0 0 0 0, clip]{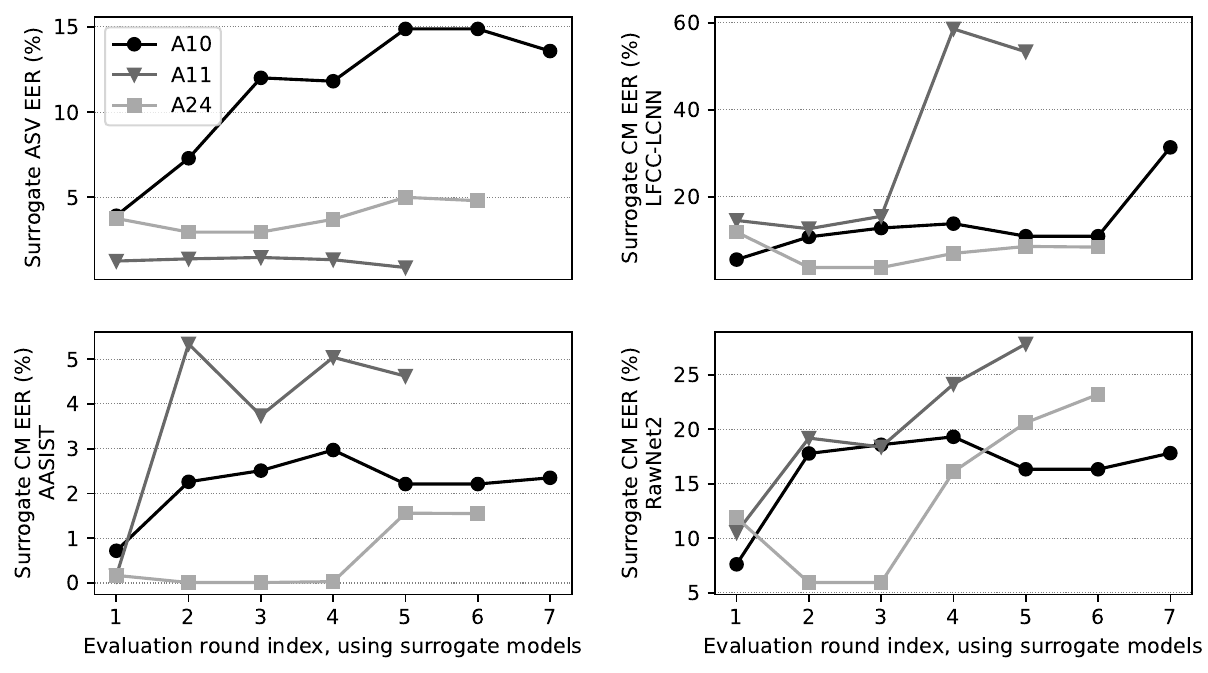}
     \caption{Progress in TTS and VC system optimisation, measured using the EERs of the surrogate models.}
     \label{fig:surrogate}
\end{figure}

Figure~\ref{fig:surrogate} illustrates the increase in ASV and CM equal error rates (EERs) which three data contributors (of five in total) were able to achieve using ASV and CM surrogates.
The four plots show EERs for each of the four surrogate models (1 ASV and 3 CM systems) and for a selection of three attacks \{A10, A11, A24\} the performance of which was evaluated using the surrogate models in multiple rounds (horizontal axis). The contributors decided the number of rounds and received results for their own attacks after each round.
A10 contributors monitored the training process and selected the checkpoint that achieved the highest ASV EER. A11 contributors selected the checkpoint that achieved the highest overall surrogate CM EER.  A24 contributors manually tweaked the training configuration of the speaker encoder (e.g.\ from independent training to joint training with the VC system) and selected the checkpoint that gave the highest surrogate ASV EER.

Results show that the surrogate models can be used successfully to implement stronger attacks. 
Use of surrogate systems among the data providers was nonetheless modest.  This is likely due to the associated optimisation cost. It is not necessarily evident how TTS and VC systems should be modified, especially given that most systems are designed to maximise some form of quality-based measure instead of EERs.  Still, some contributors were successful and it would be unwise to assume that an adversary will never optimise an attack in similar fashion. 

\begin{table}[tb!]
    \centering
    \caption{Evaluation metrics. Note that, in ASV EER, there is only one negative class of data, which can be either bona fide non-target or spoofed. In a-DCF, the bona fide non-target and spoofed data are treated as two independent negative classes. MOS EER is computed in a similar manner to CM EER but based on the output of the MOS estimator.}
    \begin{tabular}{rlll}
        \toprule
         Metric & Input scores &  Positive class & Negative class(es)  \\  \midrule
        ASV EER & ASV scores  & Bona fide AND target & (Bona fide AND non-target) OR Spoofed  \\
        CM EER &   CM scores & Bona fide & Spoofed \\ 
        a-DCF & SASV scores & Bona fide AND target & Bona fide AND non-target,  Spoofed  \\
        MOS EER & MOS & Bona fide & Spoofed \\
        \bottomrule
    \end{tabular}
    \label{tab: metric}
\end{table}

\section{Experimental validation}

\label{sec:experiment}
We describe ASV, CM, and SASV systems, experiments and results.  These serve two purposes, namely to validate the protocol design and database collection and to provide baselines for the ASVspoof~5 challenge; a comparison of different baselines is not the objective. The systems, all open-sourced, include one ASV system, a pair of CM systems, and a pair of SASV systems. As illustrated in Figure~\ref{fig:cm_sasv} and described in  \S~\ref{sec:database:challenge}, CM systems can be used on their own as solutions to Challenge Track~1. 
Solutions to Track~2 involve SASV systems, implemented either as a fusion of standalone CM and ASV subsystems or more `end-to-end' approaches (as discussed in \S~\ref{sec:database:challenge}).
Also used for validation is an open-sourced mean-opinion-score (MOS) estimator which was used to predict the perceptual quality of bona fide and differently-generated spoofed/deepfake utterances.

\subsection{Baselines, MOS estimator, and metrics}
\label{sec:baselines}

The baseline ASV system is used to gauge the threat of each attack in terms of spoofing target speakers.
It is based upon another implementation of ECAPA-TDNN~\cite{desplanques2020ecapa}\footnote{\url{https://github.com/TaoRuijie/ECAPA-TDNN}} speaker encoder, with a cosine scoring backend.
The speaker encoder is trained on VoxCeleb~2~\cite{voxceleb2} development set, with data augmentation based upon additive noises~\cite{musan2015}, room reverberation~\cite{Ko2017ASO}, and masking of input spectral features ~\cite{park19e_interspeech}.
There are two CM baselines, RawNet2~\cite{Jung2020,tak2021end} (B01) and AASIST~\cite{jung2022aasist} (B02). Both systems are trained in end-to-end fashion, as illustrated at the left to Figure~\ref{fig:cm_sasv}. These models operate directly on raw waveforms and require a fixed length 4-second audio input.
RawNet2 uses a fixed bank of $20$ sinc filters~\cite{ravanelli2018speaker} and six residual blocks with gated recurrent units (GRUs) to convert frame-level representations into utterance-level representations. Output scores are generated using fully connected layers. AASIST uses a RawNet2-based encoder~\cite{Jung2020} to extract spectro-temporal features from the raw input waveform. Spectro-temporal heterogeneous graph attention layers and max graph operations are then used to integrate temporal and spectral representations. Output scores are generated using a readout operation and a fully connected output layer.

The SASV systems are an ASV-CM fusion-based system~\cite{sasv2022} (B03, middle panel of Figure~\ref{fig:cm_sasv}), and an end-to-end system (B4~\cite{mun2023towards}, right panel of Figure~\ref{fig:cm_sasv}).
B03 is the SASV 2022 challenge baseline~\cite{sasv2022,shimbaseline} and is an LLR-based fusion~\cite{wangRevisiting2024} of the ASV and AASIST CM baselines described above.
B04, based on MFA-Conformer~\cite{zhang2022mfa}, extracts a single embedding from the input waveform and produces a single output score. It is trained in three stages: pre-training for speaker classification; copy synthesis training~\cite{wang2023spoofed} with adapted SASV loss functions; in-domain fine-tuning using ASVspoof~5 training data.

A MOS predictor~\cite{cooperGeneralizationAbilityMOS2022} is used to estimate the quality of bona fide and spoofed utterances as might be measured from human listening tests. 
It produces a MOS on a five-point scale for each input utterance. 
The system architecture is the combination of a self-supervised-learning-based wav2vec~2.0~\cite{Baevski2020} front-end, a global-average pooling layer, and a linear output layer. We used the publicly available, pre-trained system\footnote{\url{https://github.com/nii-yamagishilab/mos-finetune-ssl}} without modification. This time we did not recruit human listeners to assess the quality due to the prohibitively high manpower cost.

As depicted in Table~\ref{tab: metric}, we used the same EER metric for the evaluation of ASV and CM performance.
The ASV EER is usually estimated from a set of bona fide target trial scores (positive class) and bona fide non-target trial scores (negative class). When subjected to attacks, the latter are replaced by spoofed target scores. 

The CM EER is estimated using a set of bona fide and spoofed trial scores. SASV performance estimates are expressed in terms of the architecture-agnostic detection cost function (a-DCF)~\cite{shim2024dcf,ASVspoof5_evalplan_phase2}, a cost-based performance metric designed  specifically for the SASV task. 
While the SASV makes a binary decision, there are now three classes, namely trials involving bona fide targets (positive class) and then bona fide non-targets and spoofed target trials (negative class).
Lower a-DCF values indicate better performance.\footnote{Implementations of all baseline systems and evaluation metrics are accessible from the \href{https://github.com/asvspoof-challenge/asvspoof5}{ASVspoof~5 Github repository}.} Further details for the evaluation metrics are available in the ASVspoof~5 challenge evaluation plan~\cite{ASVspoof5_evalplan_phase2}.

The MOS EER is computed in similar fashion to the CM EER but using predicted MOS values for bona fide and spoofed utterances. Higher MOS EERs indicate spoofed utterances of (predicted) perceptual quality more comparable to that of bona fide utterances, 
while lower MOS EERs suggest that spoofed utterances are perceptually inferior to, and more easily distinguishable from bona fide utterances. We report the MOS EERs rather than the attack-level MOS mean values because the former gauges the overlap of the spoofed and bona fide MOS score distributions. %
One practical benefit is that, unlike the predicted MOS values (that may exhibit systematic domain shift), EER is invariant to any order-preserving transforms of the predicted MOS scores (including global scaling and shifting)~\cite[\S~3.1]{van2007introduction}.

\begin{table}[t!]
\caption{Performance of ASV and CM baselines in terms of EER (\%) and SASV baselines in terms of a-DCF on development set (A09-A16) and evaluation set (A17-A32). Results are shown for ASV (ECAPA-TDNN), CM baselines: B01 (RawNet2) and B02 (AASIST), SASV baselines: B03 (Fusion-based) and  B04 (single SASV system). Pooled EER and min a-DCF are computed from pooled scores across all attacks.
EERs based on the MOS scores from the MOS estimator are listed in the rightmost column. It is computed from the MOS for bona fide and spoofed utterances and hence reflects classification performance when performed using estimated quality scores alone.
A darker cell colour indicates a higher EER or min a-DCF value. ASV EERs discriminating bona fide data of target and non-target speakers are listed in the rows marked by \emph{Non-target}.
}
\setlength{\tabcolsep}{10pt}
\centering
\small
\begin{tabular}{cccccccc|c}
\toprule
 & & & \multirow{2}{*}{\shortstack{ASV \\ (EER \%)}} &  \multicolumn{2}{c}{Track 1 (EER \%)} & \multicolumn{2}{c}{Track 2 (min a-DCF)} & \multirow{2}{*}{\shortstack{MOS \\ (EER \%)}}  \\ 
\cmidrule(lr){5-6}\cmidrule(lr){7-8}
&& Attacks     &   &   {B01}     & B02    & {B03}    & \multicolumn{1}{c}{B04} &    \\ 
\midrule
\multirow{10}{*}{\rotatebox{90}{Development set}}
& & Non-target  & \cellcolor[rgb]{1.00, 1.00, 1.00} 1.88 \\ 
\cmidrule(l){3-9}
  & &      A09      & \cellcolor[rgb]{0.92, 0.92, 0.92} 16.97 & \cellcolor[rgb]{0.95, 0.95, 0.95} 16.79 & \cellcolor[rgb]{0.98, 0.98, 0.98} 7.14 & \cellcolor[rgb]{0.87, 0.87, 0.87} 0.4016 & \cellcolor[rgb]{0.95, 0.95, 0.95} 0.2009 & \cellcolor[rgb]{0.88, 0.88, 0.88} 23.78\\ 
  & &      A10      & \cellcolor[rgb]{0.92, 0.92, 0.92} 16.96 & \cellcolor[rgb]{0.95, 0.95, 0.95} 16.95 & \cellcolor[rgb]{0.98, 0.98, 0.98} 7.15 & \cellcolor[rgb]{0.87, 0.87, 0.87} 0.4027 & \cellcolor[rgb]{0.95, 0.95, 0.95} 0.1997 & \cellcolor[rgb]{0.89, 0.89, 0.89} 22.53\\ 
  & &      A11      & \cellcolor[rgb]{1.00, 1.00, 1.00} 2.13 & \cellcolor[rgb]{0.96, 0.96, 0.96} 15.17 & \cellcolor[rgb]{0.99, 0.99, 0.99} 4.81 & \cellcolor[rgb]{0.99, 0.99, 0.99} 0.0494 & \cellcolor[rgb]{0.99, 0.99, 0.99} 0.0272 & \cellcolor[rgb]{0.94, 0.94, 0.94} 13.55\\ 
  & &      A12      & \cellcolor[rgb]{0.75, 0.75, 0.75} 37.10 & \cellcolor[rgb]{0.59, 0.59, 0.59} 78.86 & \cellcolor[rgb]{0.59, 0.59, 0.59} 78.90 & \cellcolor[rgb]{0.61, 0.61, 0.61} 0.8465 & \cellcolor[rgb]{0.65, 0.65, 0.65} 0.8549 & \cellcolor[rgb]{0.99, 0.99, 0.99} 1.82\\ 
  & &      A13      & \cellcolor[rgb]{1.00, 1.00, 1.00} 1.49 & \cellcolor[rgb]{0.85, 0.85, 0.85} 40.18 & \cellcolor[rgb]{0.96, 0.96, 0.96} 14.43 & \cellcolor[rgb]{0.99, 0.99, 0.99} 0.0353 & \cellcolor[rgb]{1.00, 1.00, 1.00} 0.0216 & \cellcolor[rgb]{0.97, 0.97, 0.97} 7.81\\ 
  & &      A14      & \cellcolor[rgb]{1.00, 1.00, 1.00} 1.47 & \cellcolor[rgb]{0.94, 0.94, 0.94} 21.43 & \cellcolor[rgb]{1.00, 1.00, 1.00} 1.18 & \cellcolor[rgb]{0.99, 0.99, 0.99} 0.0333 & \cellcolor[rgb]{1.00, 1.00, 1.00} 0.0221 & \cellcolor[rgb]{1.00, 1.00, 1.00} 0.07\\ 
  & &      A15      & \cellcolor[rgb]{0.98, 0.98, 0.98} 6.20 & \cellcolor[rgb]{0.92, 0.92, 0.92} 24.03 & \cellcolor[rgb]{0.96, 0.96, 0.96} 12.93 & \cellcolor[rgb]{0.96, 0.96, 0.96} 0.1449 & \cellcolor[rgb]{0.98, 0.98, 0.98} 0.0746 & \cellcolor[rgb]{0.91, 0.91, 0.91} 18.70\\ 
  & &      A16      & \cellcolor[rgb]{0.98, 0.98, 0.98} 5.90 & \cellcolor[rgb]{0.88, 0.88, 0.88} 33.34 & \cellcolor[rgb]{0.94, 0.94, 0.94} 19.31 & \cellcolor[rgb]{0.96, 0.96, 0.96} 0.1422 & \cellcolor[rgb]{0.99, 0.99, 0.99} 0.0662 & \cellcolor[rgb]{0.97, 0.97, 0.97} 6.69\\ 
  \cmidrule{3-9}
  & &  Pooled   & \cellcolor[rgb]{0.94, 0.94, 0.94} 14.66 & \cellcolor[rgb]{0.90, 0.90, 0.90} 29.49 & \cellcolor[rgb]{0.95, 0.95, 0.95} 17.83 & \cellcolor[rgb]{0.90, 0.90, 0.90} 0.3156 & \cellcolor[rgb]{0.94, 0.94, 0.94} 0.2254 & \cellcolor[rgb]{0.94, 0.94, 0.94} 14.31\\ 
\midrule
\multirow{18}{*}{\rotatebox{90}{Evaluation set}}
& & Non-target & \cellcolor[rgb]{0.98, 0.98, 0.98} 5.22 & \\ 
\cmidrule(l){3-9}
  & \multirow{9}{*}{\rotatebox{90}{TTS/VC }}
  &      A17      & \cellcolor[rgb]{0.65, 0.65, 0.65} 44.64 & \cellcolor[rgb]{0.93, 0.93, 0.93} 22.58 & \cellcolor[rgb]{0.95, 0.95, 0.95} 16.44 & \cellcolor[rgb]{0.59, 0.59, 0.59} 0.8818 & \cellcolor[rgb]{0.81, 0.81, 0.81} 0.5598 & \cellcolor[rgb]{0.73, 0.73, 0.73} 43.07\\ 
  &&      A19      & \cellcolor[rgb]{0.59, 0.59, 0.59} 49.80 & \cellcolor[rgb]{0.71, 0.71, 0.71} 63.75 & \cellcolor[rgb]{0.73, 0.73, 0.73} 59.99 & \cellcolor[rgb]{0.59, 0.59, 0.59} 0.8881 & \cellcolor[rgb]{0.59, 0.59, 0.59} 0.9450 & \cellcolor[rgb]{0.95, 0.95, 0.95} 13.05\\ 
  &&      A21      & \cellcolor[rgb]{0.81, 0.81, 0.81} 30.74 & \cellcolor[rgb]{0.92, 0.92, 0.92} 25.67 & \cellcolor[rgb]{0.95, 0.95, 0.95} 17.05 & \cellcolor[rgb]{0.71, 0.71, 0.71} 0.7187 & \cellcolor[rgb]{0.89, 0.89, 0.89} 0.3647 & \cellcolor[rgb]{0.78, 0.78, 0.78} 37.27\\ 
  &&      A22      & \cellcolor[rgb]{0.84, 0.84, 0.84} 27.52 & \cellcolor[rgb]{0.92, 0.92, 0.92} 24.50 & \cellcolor[rgb]{0.95, 0.95, 0.95} 17.63 & \cellcolor[rgb]{0.75, 0.75, 0.75} 0.6402 & \cellcolor[rgb]{0.85, 0.85, 0.85} 0.4731 & \cellcolor[rgb]{0.84, 0.84, 0.84} 30.05\\ 
  &&      A24      & \cellcolor[rgb]{0.89, 0.89, 0.89} 21.08 & \cellcolor[rgb]{0.92, 0.92, 0.92} 23.61 & \cellcolor[rgb]{0.96, 0.96, 0.96} 13.35 & \cellcolor[rgb]{0.83, 0.83, 0.83} 0.4846 & \cellcolor[rgb]{0.90, 0.90, 0.90} 0.3363 & \cellcolor[rgb]{0.81, 0.81, 0.81} 33.10\\ 
  &&      A25      & \cellcolor[rgb]{0.96, 0.96, 0.96} 9.96 & \cellcolor[rgb]{0.90, 0.90, 0.90} 29.78 & \cellcolor[rgb]{0.94, 0.94, 0.94} 21.01 & \cellcolor[rgb]{0.94, 0.94, 0.94} 0.2373 & \cellcolor[rgb]{0.90, 0.90, 0.90} 0.3325 & \cellcolor[rgb]{0.96, 0.96, 0.96} 9.15\\ 
  &&      A26      & \cellcolor[rgb]{0.93, 0.93, 0.93} 15.77 & \cellcolor[rgb]{0.84, 0.84, 0.84} 41.95 & \cellcolor[rgb]{0.89, 0.89, 0.89} 31.35 & \cellcolor[rgb]{0.88, 0.88, 0.88} 0.3765 & \cellcolor[rgb]{0.85, 0.85, 0.85} 0.4781 & \cellcolor[rgb]{0.90, 0.90, 0.90} 20.86\\ 
  &&      A28      & \cellcolor[rgb]{0.73, 0.73, 0.73} 38.44 & \cellcolor[rgb]{0.85, 0.85, 0.85} 39.43 & \cellcolor[rgb]{0.88, 0.88, 0.88} 32.10 & \cellcolor[rgb]{0.61, 0.61, 0.61} 0.8519 & \cellcolor[rgb]{0.67, 0.67, 0.67} 0.8133 & \cellcolor[rgb]{0.59, 0.59, 0.59} 55.96\\ 
  &&      A29      & \cellcolor[rgb]{0.80, 0.80, 0.80} 31.88 & \cellcolor[rgb]{0.95, 0.95, 0.95} 17.54 & \cellcolor[rgb]{0.97, 0.97, 0.97} 8.93 & \cellcolor[rgb]{0.70, 0.70, 0.70} 0.7328 & \cellcolor[rgb]{0.91, 0.91, 0.91} 0.3163 & \cellcolor[rgb]{0.66, 0.66, 0.66} 49.60\\ 
  \cmidrule{3-9}
  & \multirow{7}{*}{\rotatebox{90}{Adversarial}}
  &      A18      & \cellcolor[rgb]{0.77, 0.77, 0.77} 34.28 & \cellcolor[rgb]{0.75, 0.75, 0.75} 57.64 & \cellcolor[rgb]{0.79, 0.79, 0.79} 50.02 & \cellcolor[rgb]{0.69, 0.69, 0.69} 0.7434 & \cellcolor[rgb]{0.73, 0.73, 0.73} 0.7223 & \cellcolor[rgb]{0.96, 0.96, 0.96} 9.16\\ 
  &&      A20      & \cellcolor[rgb]{0.76, 0.76, 0.76} 35.82 & \cellcolor[rgb]{0.81, 0.81, 0.81} 46.95 & \cellcolor[rgb]{0.88, 0.88, 0.88} 33.70 & \cellcolor[rgb]{0.70, 0.70, 0.70} 0.7244 & \cellcolor[rgb]{0.63, 0.63, 0.63} 0.8794 & \cellcolor[rgb]{0.99, 0.99, 0.99} 1.34\\ 
  &&      A23      & \cellcolor[rgb]{0.84, 0.84, 0.84} 27.13 & \cellcolor[rgb]{0.89, 0.89, 0.89} 30.35 & \cellcolor[rgb]{0.88, 0.88, 0.88} 32.78 & \cellcolor[rgb]{0.75, 0.75, 0.75} 0.6420 & \cellcolor[rgb]{0.84, 0.84, 0.84} 0.5015 & \cellcolor[rgb]{0.95, 0.95, 0.95} 12.97\\ 
  &&      A27      & \cellcolor[rgb]{0.77, 0.77, 0.77} 34.68 & \cellcolor[rgb]{0.85, 0.85, 0.85} 38.85 & \cellcolor[rgb]{0.91, 0.91, 0.91} 27.60 & \cellcolor[rgb]{0.72, 0.72, 0.72} 0.6994 & \cellcolor[rgb]{0.81, 0.81, 0.81} 0.5662 & \cellcolor[rgb]{1.00, 1.00, 1.00} 1.10\\ 
  &&      A30      & \cellcolor[rgb]{0.70, 0.70, 0.70} 41.12 & \cellcolor[rgb]{0.84, 0.84, 0.84} 42.26 & \cellcolor[rgb]{0.83, 0.83, 0.83} 42.85 & \cellcolor[rgb]{0.66, 0.66, 0.66} 0.7798 & \cellcolor[rgb]{0.76, 0.76, 0.76} 0.6737 & \cellcolor[rgb]{0.99, 0.99, 0.99} 3.37\\ 
  &&      A31      & \cellcolor[rgb]{0.75, 0.75, 0.75} 37.14 & \cellcolor[rgb]{0.88, 0.88, 0.88} 33.22 & \cellcolor[rgb]{0.91, 0.91, 0.91} 27.48 & \cellcolor[rgb]{0.70, 0.70, 0.70} 0.7220 & \cellcolor[rgb]{0.81, 0.81, 0.81} 0.5641 & \cellcolor[rgb]{0.98, 0.98, 0.98} 5.31\\ 
  &&      A32      & \cellcolor[rgb]{0.82, 0.82, 0.82} 29.30 & \cellcolor[rgb]{0.90, 0.90, 0.90} 29.78 & \cellcolor[rgb]{0.94, 0.94, 0.94} 19.50 & \cellcolor[rgb]{0.76, 0.76, 0.76} 0.6299 & \cellcolor[rgb]{0.82, 0.82, 0.82} 0.5481 & \cellcolor[rgb]{1.00, 1.00, 1.00} 0.80\\ 
  \cmidrule{3-9}
  &&  Pooled  & \cellcolor[rgb]{0.79, 0.79, 0.79} 32.11 & \cellcolor[rgb]{0.87, 0.87, 0.87} 36.04 & \cellcolor[rgb]{0.90, 0.90, 0.90} 29.12 & \cellcolor[rgb]{0.73, 0.73, 0.73} 0.6806 & \cellcolor[rgb]{0.81, 0.81, 0.81} 0.5741 & \cellcolor[rgb]{0.86, 0.86, 0.86} 26.55\\ 
\bottomrule
\end{tabular}
\label{tab: baseline results}
\end{table}

\begin{figure}[!t]
    \centering
    \includegraphics[width=0.9\linewidth]{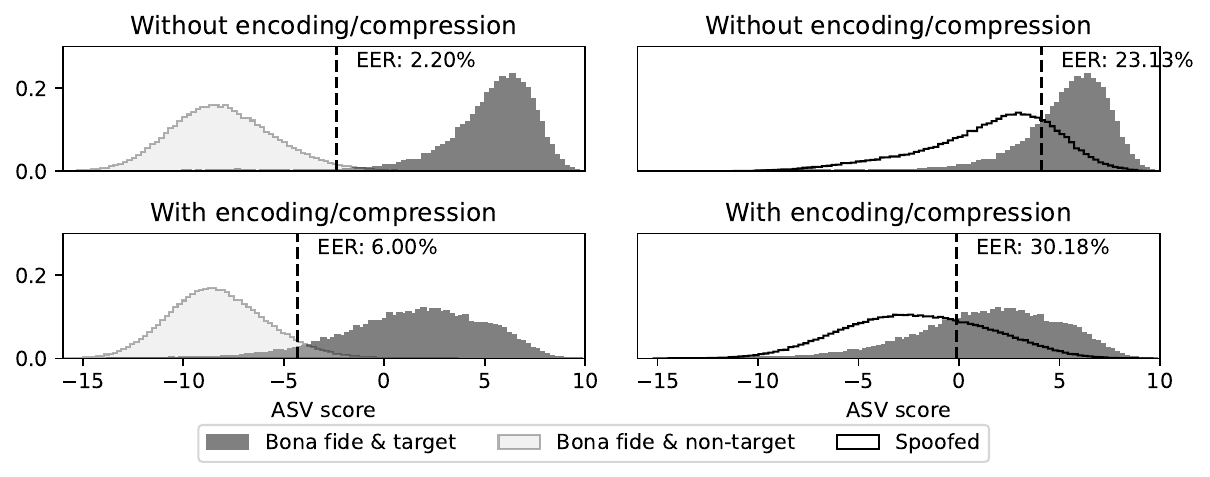}
    \vspace{-2mm}
    \caption{Distribution of ASV scores on the evaluation set when the trials are processed with encoding/compression (top-row) or without (bottom-row). Sub-figures in the left column show the distributions for bona fide utterances from target and non-target speakers. Sub-figures in the right column show the distributions for bona fide utterances from target speakers and spoofed utterances from all the attacks. The thresholds producing the EERs are plotted as vertical dash lines.}
    \label{fig:asv_scores}
\end{figure}

\subsection{Impacts upon ASV}
\label{sec:experiment:asv}
In the following we report an evaluation of the baseline ASV system when it encounters bona fide target and bona fide non-target trials only, and then when it is subject to attack.  
We report EERs only in the following, readers with an interest in performance at other operating points are referred to detection error tradeoff (DET) profiles~\cite{martin1997det} presented in \ref{app:det}. 

ASV results for the development and evaluation sets are listed in the second column of Table~\ref{tab: baseline results}. 
When assessed using a mix of only target and non-target trials the EER is 1.88\% for the development set and 5.22\% for the evaluation set.  The higher EER for the evaluation set is due mostly to the variation in encoding/compression and bandwidth. The effect of encoding/compression is illustrated in the score histograms shown in the left column on Figure~\ref{fig:asv_scores}.  They show score distributions without (top) and with (middle) encoding/compression, along with their EERs of 2.20\% and 6.00\%, respectively.
Encoding/compression reduces the differences between the distributions resulting in a larger overlap and therefore a higher EER.

Also shown in the second column of Table~\ref{tab: baseline results} are EERs for the same ASV system when evaluated using a mix of target and spoofed/deepfake trials.  When the ASV system is under attack, EERs are almost universally higher, indicating that most attacks are successful in provoking higher scores.
A11, A13, and A14 attacks, all in the development set, are relatively ineffective and lead to a trivial increase or even a reduction in the EER. This finding is consistent with observations made from the t-SNE visualisations shown in Figure~\ref{fig:tsne} wherein the same three attacks are among the outliers and far from the centroid of the cluster for bona fide utterances.  
A12, a simplified unit-selection based TTS system, generates the highest EER of 37.10\% for the development set. 

Attacks in the evaluation set result in some of the highest EERs.
Of the 16 attacks, 14 lead to EERs higher than 20\%. 
A19, also a legacy, unit-selection based TTS system, is once again the attack which provokes the highest EER. 
These results are in line with previously reported findings~\cite{asvspoof2019database, jung_what_2024}, a challenge recognized since the very first ASVspoof challenge~\cite{Wu-ASVspoof2015,Wu2017-ASVspoof-IEEE-J-STSP}. A decade after, this persistent challenge is worthwhile emphasising: \emph{the most effective attack algorithms are not necessarily those implemented with the latest technology}. Legacy TTS systems, based on the concatenation of bona fide speech segments extracted from the utterances of a target speaker are particularly effective.

Pre-trained TTS attacks A17, A28, and A29 give the next highest EERs, all above 31\%.\footnote{A small portion of A17's training data consists of utterances from 12 target speakers but does not overlapped with the ASVspoof~5 evaluation set (Section~\ref{sec:attacks:eval}). The ASV EER computed on the bona fide and spoofed data of the `seen' 12 target speakers is 42.51\%, which is lower than that of the `unseen' (44.65\%). The high ASV EER provoked by A17 is not due to the `seen' target speakers.} 
A21 and A22, both trained using only data specified in the ASVspoof~5 contributors' protocol, also provoke high EERs. %
VC-based attacks A24, A25 and A26 are among the least effective.
The effect of encoding/compression for the evaluation set is illustrated to the right in Figure~\ref{fig:asv_scores}, which shows score distributions for bona fide target and spoofed/deepfake trials. As one might expect, once again, encoding/compression reduces the differences between the distributions, provoking %
higher EERs.

Results for the set of adversarial attacks A18, A20, A23, A27, A30, A31, and A32 shown to the bottom of Table~\ref{tab: baseline results} %
also cause a significant degradation in ASV performance. Malacopula adversarial attacks, designed to compromise ASV systems, result in higher EERs than their underlying TTS/VC attacks (A26 15.77\% $\rightarrow$ A27 34.68\%, A22 27.52\% $\rightarrow$ A31 37.14\%, A25 9.96\% $\rightarrow$ A32 29.30\%).  Malafide attacks, which are not designed to compromise ASV systems, are ineffective with one exception.  
For A23, the EER increases to 27.13\% from the 16.97\% EER for the underlying A09 attack.

In the following we seek to determine if the attacks might also be more difficult to detect using a CM designed, unlike ASV systems, specifically for this task.

\subsection{CM results}
Table~\ref{tab: baseline results} shows performance of both CM baselines, B01 and B02, in terms of the EER (\%). %
For the development set, the unit-selection based TTS attack A12 is the most challenging to detect.\footnote{A12 provokes EERs higher than 50\%. An EER higher than 50\% means that the majority of CM scores for spoofed/deepfake utterances are higher than those for bona fide utterances. By tradition, higher CM scores indicate greater support for the bona fide hypothesis. Therefore, from the perspective the baselines CMs, the spoofed/deepfake utterance of A12 are more likely to be `bona fide' than the actual bona fide utterances.}
Even though the t-SNE plot in Figure~\ref{fig:tsne} shows that A13 is an outlier, it is the second most difficult to detect for B01, while the EER for B02 is much lower.  For B02 the second most difficult to detect is A16, a zero-shot VC system, for which the EER is also high for B01.  All other attacks in the development set produce lower EERs for each CM, but many are still high.  Pooled EERs of 29.49\% and 17.83\% show that neither CM is especially reliable in detecting the full set of attacks.

EERs for attacks in the evaluation set are generally higher, increasing to 36.04\% and 29.12\%.  
Higher EERs are expected given that the evaluation set includes more advanced attack algorithms as well as a large portion of encoded/compressed data. Interested readers can refer to Table~\ref{tab:baseline_codec} for a breakdown of results with and without encoding/compression. Those shown in Table~\ref{tab: baseline results} are for pooled results derived using the ASVspoof~5 evaluation set.

Last, except for attack A20, Malafide adversarial filtering, which is applied to already-spoofed/deepfake utterances to further compromise the CM, leads to additional increases in the EER. As shown in Table~\ref{tab: baseline results}, Malafide filtering provokes increases in the EER in the cases of A17 \{22.58\%, 16.44\%\} $\rightarrow$ A18 \{57.64\%, 50.02\%\}, A09 \{16.79\%, 7.14\%\} $\rightarrow$ A23 \{30.35\%, 32.78\%\}.
The comparison of results for A17 and A30 shows that the combination of Malafide and Malacopula filtering also provokes increases in the EER. However, the application of Malacopula alone (designed to further compromise an ASV system), generally fails to increase the CM EER.  
Only for A31 does Malacopula filtering increase the EER beyond that for the underlying TTS attack A22.

\subsection{SASV results}
Thus far we have seen that while some attacks, e.g.\ A13 and A14, fail to provoke high ASV EERs, they can be difficult to detect, especially in the case of baseline B01.  For other attacks, e.g.\ A12, ASV and CM EERs are both high, while for others they are both relatively low, e.g.\ for A14 and baseline B02.  For all of these cases there are separate ASV and CM sub-systems.  In the following we explore the reliability of SASV systems which produce a single score. Results are presented for B03 and B04 SASV systems in Table~\ref{tab: baseline results} in terms of the min a-DCF~\cite{shim2024dcf}. %
Again, lower a-DCF values indicate better performance. %
Results indicate that both SASV baselines struggle with unit-selection based TTS attacks A12 and A19. 
For both attacks and both baselines, the min a-DCF values are the highest for the development and evaluation sets.
The A28 attack, based upon the recent YourTTS system, also produces high min a-DCF values, again for both baselines, indicating the higher threat of more sophisticated attacks. 

For both the development and evaluation sets, attacks that provoke the lowest increase in ASV EERs, e.g.\ A25 and A26, also provoke the lowest min a-DCF values, even when CM EERs are relatively high. This implies that the B03 baseline, which is based upon the score-level fusion of independent CM and ASV systems, is reasonably successful in combining the merits of each in protecting against spoofs/deepfakes.
Malacopula attacks A27, A31, and A32 result in higher min a-DCF values. In contrast, for B03, Malafide attacks A18, A20, and A23 fail to increase min a-DCF values beyond those of the underlying attacks.

Even so, pooled min a-DCF values remain high for both baselines and are substantially higher for the evaluation set than for the development set: pooled min a-DCF values for the development set are 0.3156 and 0.2254, while those for the evaluation subset are 0.6806  and 0.5741. %
A similar trend is observed under encoding/compression conditions (See Table~\ref{tab:baseline_codec}), where the a-DCF values on the evaluation set are even higher. %
These results indicate the challenge to maintain reliability when faced with the most advanced attacks and in the condition of encoding/compression. We would also like to remind the reader that the \emph{minimum} a-DCF provides \emph{the most optimistic} view of evaluation set performance. In particular, similar to EER, the minimum a-DCF metric involves an `oracle' detection threshold obtained using ground truth labels. While useful for analysis of results, such as those presented in Table~\ref{tab: baseline results}, the ground truth labels are \emph{never available} in real-world operational settings (otherwise, why bother about developing classifiers?). The threshold(s) must be selected in advance at the system development time. Substantial future challenge remains on this front, particularly when we are faced with novel attacks and coded/compressed data. Investigation of the decision threshold is left to an upcoming analysis paper on the ASVspoof~5 challenge results.

\subsection{MOS results}
\label{sec:experiment:mos}

Estimated MOS EERs for each attack are shown in the rightmost column of Table~\ref{tab: baseline results}. Lower MOS EERs suggest that spoofed data is of perceptually lower quality than bona fide data.
As one might have expected, the results indicate that spoofed/deepfake utterances of lower (estimated) perceptual quality are generally easier to detect, whereas those of higher quality present a greater challenge. 
Pre-trained TTS system A28 produces the highest of all MOS EERs. The same attack also provokes a high ASV EER, high CM EER, and high SASV min a-DCF value. Other TTS/VC attacks, namely, A17, A21, A22, A24, and A29, also lead to poor performance of baselines in the case of all metrics, even if A29 is detected by B02 with an CM EER below 10\%.

There are some exceptions, however. Some attacks with a low MOS EER nonetheless pose a substantial threat. 
For the legacy unit-selection TTS system A12 the MOS EER is less than 2\%, though it provokes CM EERs higher than 78\%. This finding suggests that, while human listeners may be able to distinguish easily between  bona fide utterances and spoofs/deepfakes generated using the A12 attack, the challenge for automatic systems is substantial. We observe similar results for A19 and the full set of adversarial attacks.  
Casual listening test reveal notable concatenation artefacts in the case of A12 and A19, and the comparatively poor quality of adversarial attacks. 
The quality of speech produced by concatenative TTS approaches is limited by the number of speech units available among adaptation utterances (\S~\ref{sec:attacks:eval}) while the filter coefficients of the two adversarial attacks are optimised to %
misguide only automatic systems---but not to preserve perceived quality~\cite{malacopula}. The combination of quality-based estimates with more traditional CM techniques may lead to more generalisable and reliable detection techniques.

\section{Conclusions}

We described the design, collection and validation of the ASVspoof~5 database.
Adoption of a new source database, itself collected by volunteers in the wild, gives rise to new challenges, involving both the generation of spoofed and deepfake speech as well as in detection.
The ASVspoof~5 database contains greater acoustic diversity and speech data collected from a vastly greater number of speakers than any previous ASVspoof database.
Spoofed/deepfake speech, generated with a mix of legacy and contemporary text-to-speech synthesis, voice conversion and adversarial attack technology, is also crowdsourced, further adding to variability.
The ASVspoof~5 database supports the evaluation of automatic speaker verification systems when subjected to attacks, of independent spoofing/deepfake detection solutions, and of spoofing-robust automatic speaker verification systems of almost any architecture.

Comprehensive database validation experiments and results
show the challenges to develop  robust, generalisable detection, especially when faced with data encoded and compressed using the latest neural codecs.
Despite the use of more challenging source data, both legacy and contemporary generative technology still pose a grave threat to the reliability of automatic systems, and show the need for continued progress in spoofed/deepfake detection.

The protocols and other resources described in this paper will be helpful to the participants of the ASVspoof~5 challenge who wish to analyse their results, as we hope it will be to others seeking to use the database to support other work in spoofing/deepfake detection.  Most of the resources described in the paper are all already freely available to the community and can be used to reproduce our results.
Being mindful of the obvious ethical considerations and potential for their misuse, the generation protocols and access to surrogate systems are available only upon request.  They should be used to help the spoofing/deepfake detection community track future developments in generative technology, e.g.\ by contributing data produced using algorithms which emerge in the future, or for other beneficial and harmless applications.

\section{Acknowledgement}

The ASVspoof~5 organising committee expresses its gratitude and appreciation to other contributors: Hengcheng Kuo and Hung-yi Lee, National Taiwan University; Yihan Wu, Renmin University of China; Yu Tsao, Academia Sinica; Minki Kang, KRAFTON, Korea.

This work is partially supported by JST, PRESTO Grant Number JPMJPR23P9, JST AIP Acceleration Project Grant Number JPMJCR24U3, Japan, and with funding received from the French Agence Nationale de la Recherche (ANR) via
the BRUEL (ANR-22-CE39-0009) and COMPROMIS (ANR22-PECY-0011) projects. This work was also partially supported by the Academy of Finland (Decision No. 349605, project ``SPEECHFAKES'').
Part of the computation and data generation was performed using the TSUBAME4.0 supercomputer at Tokyo Institute of Science, and by supercomputing infrastructure provided by CSC --- IT Center for Science (Finland).

\newpage
\appendix
\section{Encoding and compression condition C11}
\label{app:codec11}

\begin{table}[!ht]
    \centering
    \normalsize
    \caption{C11 configurations.}
    \setlength{\tabcolsep}{4pt}
    \begin{tabular}{clll}
        \toprule
        ID & Device type & Calling device / software & Audio injection method \\ 
        \midrule
        C11-1 & PC & Microsoft Teams & Virtual audio cable driver \\
        C11-2 & PC & Microsoft Teams w/ noise cancellation & Virtual audio cable driver \\
        C11-3 & Smartphone & Poco Phone 4 5G & Bluetooth \\
        C11-4 & Smartphone & Redmi Note 8 Pro & Bluetooth \\
        C11-5 & Smartphone & Redmi Note 8 Pro & Analog cable \\
        C11-6 & Smartphone & Samsung Galaxy A12 & Bluetooth \\
        C11-7 & Smartphone & Samsung Galaxy A12 & Analog wired connection \\
        C11-8 & Smartphone & Samsung Galaxy S23 Ultra & Bluetooth \\ 
        \bottomrule
    \end{tabular}
    \vspace{-3mm}
    \label{tab:C11_configs}
\end{table}

\section{Dendrogram of ASVspoof 5 data}
\label{app:dendrogram}
Figure~\ref{fig:embedding-dendrogram} is plotted with the following configurations. We first compute a $33 \times 33$ pairwise similarity matrix $\boldsymbol{S} = [s(X_i,X_j)]$ that represents pairwise similarity between collections of speaker embeddings corresponding to each attack (or bona fide utterances). This similarity is computed as
    \begin{equation}
        \begin{aligned}
        s(X_i, X_j) & = 1 - \frac{1}{2}\Big(\frac{1}{|X_i|}\sum_{\boldsymbol{a} \in X_i} \min_{\boldsymbol{b} \in X_j} c(\boldsymbol{a},\boldsymbol{b}) + \frac{1}{|X_j|}\sum_{\boldsymbol{a} \in X_j} \min_{\boldsymbol{b} \in X_i} c(\boldsymbol{a},\boldsymbol{b})\Big),
        \end{aligned}
    \end{equation}
where $|\cdot|$ denotes the number of observations and $c(\boldsymbol{a},\boldsymbol{b})$ is cosine distance of vectors $\boldsymbol{a}$ and $\boldsymbol{b}$. The similarity matrix $\boldsymbol{S}$ is used for hierarchical agglomerative clustering, using the Ward's minimum variance criterion as the clustering objective. The figure is rendered using the Seaborn \texttt{clustermap} API~\cite{Waskom2021}.

\newpage
\section{DET Curves}
\label{app:det}
\begin{figure}[h!]
    \centering
    \subfloat[Baseline ASV]{
        \includegraphics[width=0.5\textwidth, trim=7.5 0 0 0, clip]{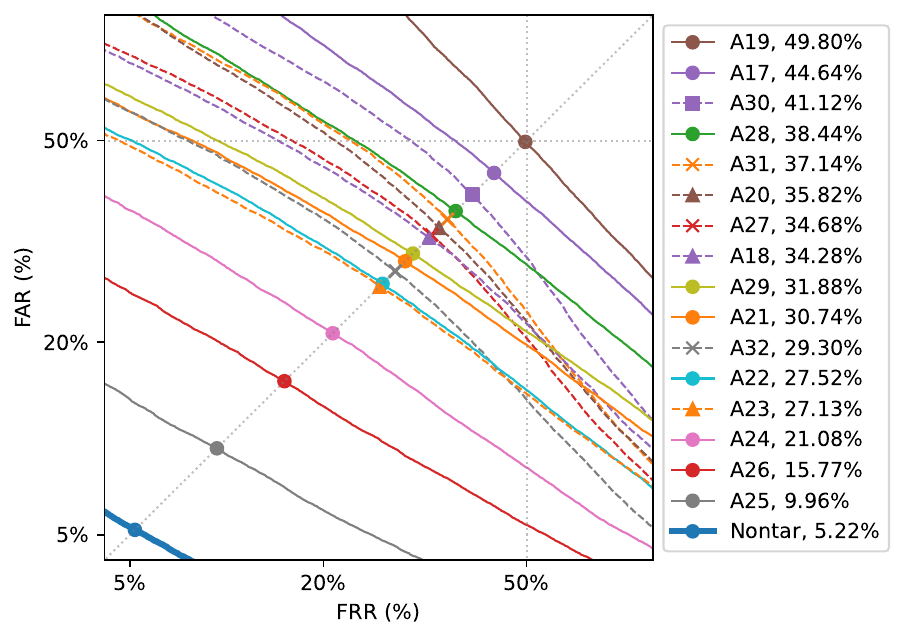}
        \label{fig:det_asv}
    }
    \subfloat[Baseline CM B01]{
        \includegraphics[width=0.5\textwidth, trim=7.7 0 0 0, clip]{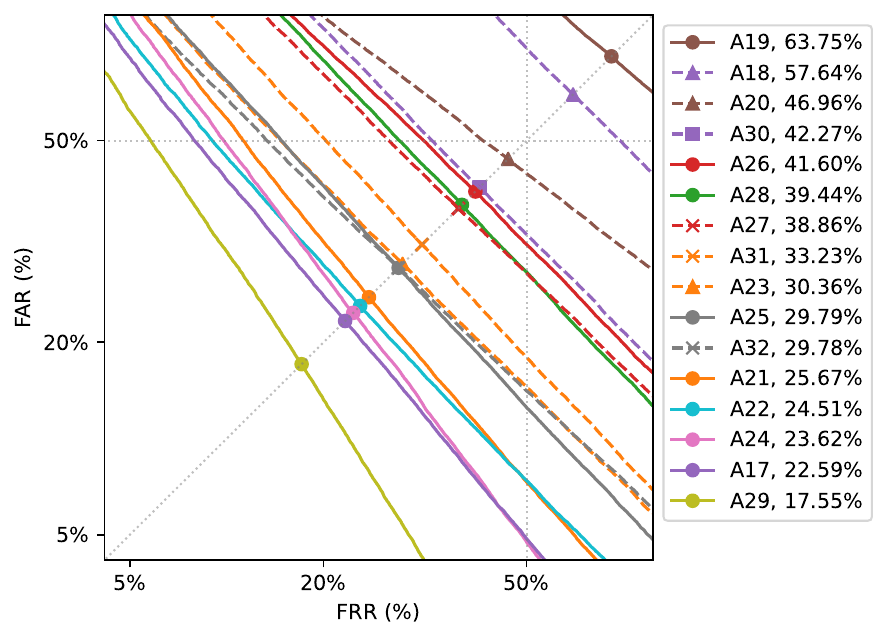}
        \label{fig:det_rawnet_cm}
    }\\
    \subfloat[Baseline CM B02]{
        \includegraphics[width=0.5\textwidth, trim=7.7 0 0 0, clip]{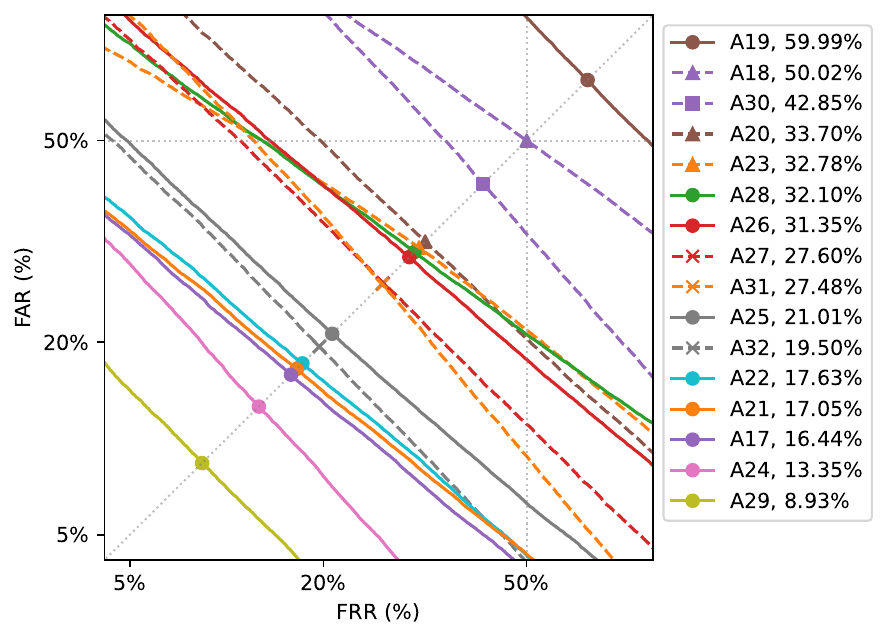}
        \label{fig:det_aasist_cm}
    }
     \caption{DET curves of ASV and CM baselines illustrating performance across different attacks in the evaluation set. Solid lines are zero-effort (from non-target speakers) and non-adversarial attacks. Dashed lines are adversarial attacks, including Malafide, Malacopula, and their combination. The colour of the dashed line is consistent with the non-adversarial attack counterpart. EER operating points of non-adversarial, Malafide, Malacopula, and combination of Malafide and Malacopula attacks are marked with $\bullet, \blacktriangle, \times$, and $\blacksquare$, respectively. Numbers listed in the legend are EER values. The order of attacks in the legend is sorted based on the EER.}
     \label{fig:det_eer}
\end{figure}

\newpage
\section{Results with and without encoding/compressiong}

\begin{table}[h!]
\caption{Performance of ASV and CM baselines in terms of EER (\%) and SASV baselines in terms of a-DCF on evaluation set (A17-A32), with and without encoding/compression. The results are presented in a similar fashion to Table~\ref{tab: baseline results} except that the upper and bottom sub-tables show the breakdown results in conditions with and without encoding/compression, respectively.
}
\setlength{\tabcolsep}{10pt}
\centering
\small
\begin{tabular}{cccccccc|c}
\toprule
 & & & \multirow{2}{*}{\shortstack{ASV \\ (EER \%)}} &  \multicolumn{2}{c}{Track 1 (EER \%)} & \multicolumn{2}{c}{Track 2 (min a-DCF)} & \multirow{2}{*}{\shortstack{MOS \\ (EER \%)}}  \\ 
\cmidrule(lr){5-6}\cmidrule(lr){7-8}
&& Attacks     &   &   {B01}     & B02    & {B03}    & \multicolumn{1}{c}{B04} &    \\ 
\midrule
\multirow{18}{*}{\rotatebox{90}{Evaluation set \textbf{without} encoding/compression}}
& & Non-target & \cellcolor[rgb]{1.00, 1.00, 1.00} 2.20 & \\ 
\cmidrule(l){3-9}
  & \multirow{9}{*}{\rotatebox{90}{TTS/VC }}
      &  A17   & \cellcolor[rgb]{0.76, 0.76, 0.76} 37.44 & \cellcolor[rgb]{0.95, 0.95, 0.95} 14.48 & \cellcolor[rgb]{0.97, 0.97, 0.97} 8.30 & \cellcolor[rgb]{0.62, 0.62, 0.62} 0.8486 & \cellcolor[rgb]{0.95, 0.95, 0.95} 0.2032 & \cellcolor[rgb]{0.90, 0.90, 0.90} 23.02\\ 
  &   &  A19   & \cellcolor[rgb]{0.59, 0.59, 0.59} 52.53 & \cellcolor[rgb]{0.59, 0.59, 0.59} 66.46 & \cellcolor[rgb]{0.59, 0.59, 0.59} 62.60 & \cellcolor[rgb]{0.61, 0.61, 0.61} 0.8708 & \cellcolor[rgb]{0.61, 0.61, 0.61} 0.9053 & \cellcolor[rgb]{0.99, 0.99, 0.99} 3.34\\ 
  &   &  A21   & \cellcolor[rgb]{0.93, 0.93, 0.93} 16.19 & \cellcolor[rgb]{0.94, 0.94, 0.94} 17.27 & \cellcolor[rgb]{0.98, 0.98, 0.98} 6.63 & \cellcolor[rgb]{0.88, 0.88, 0.88} 0.3873 & \cellcolor[rgb]{0.97, 0.97, 0.97} 0.1060 & \cellcolor[rgb]{0.93, 0.93, 0.93} 17.94\\ 
  &   &  A22   & \cellcolor[rgb]{0.95, 0.95, 0.95} 13.72 & \cellcolor[rgb]{0.93, 0.93, 0.93} 18.72 & \cellcolor[rgb]{0.97, 0.97, 0.97} 8.05 & \cellcolor[rgb]{0.90, 0.90, 0.90} 0.3263 & \cellcolor[rgb]{0.95, 0.95, 0.95} 0.2018 & \cellcolor[rgb]{0.95, 0.95, 0.95} 13.30\\ 
  &   &  A24   & \cellcolor[rgb]{0.97, 0.97, 0.97} 9.31 & \cellcolor[rgb]{0.95, 0.95, 0.95} 15.36 & \cellcolor[rgb]{0.99, 0.99, 0.99} 3.63 & \cellcolor[rgb]{0.94, 0.94, 0.94} 0.2141 & \cellcolor[rgb]{0.97, 0.97, 0.97} 0.1065 & \cellcolor[rgb]{0.94, 0.94, 0.94} 15.11\\ 
  &   &  A25   & \cellcolor[rgb]{0.99, 0.99, 0.99} 3.51 & \cellcolor[rgb]{0.92, 0.92, 0.92} 20.48 & \cellcolor[rgb]{0.96, 0.96, 0.96} 9.76 & \cellcolor[rgb]{0.98, 0.98, 0.98} 0.0814 & \cellcolor[rgb]{0.98, 0.98, 0.98} 0.0813 & \cellcolor[rgb]{0.99, 0.99, 0.99} 1.70\\ 
  &   &  A26   & \cellcolor[rgb]{0.98, 0.98, 0.98} 6.01 & \cellcolor[rgb]{0.83, 0.83, 0.83} 36.31 & \cellcolor[rgb]{0.92, 0.92, 0.92} 20.03 & \cellcolor[rgb]{0.96, 0.96, 0.96} 0.1423 & \cellcolor[rgb]{0.96, 0.96, 0.96} 0.1571 & \cellcolor[rgb]{0.97, 0.97, 0.97} 6.79\\ 
  &   &  A28   & \cellcolor[rgb]{0.86, 0.86, 0.86} 26.67 & \cellcolor[rgb]{0.85, 0.85, 0.85} 33.42 & \cellcolor[rgb]{0.90, 0.90, 0.90} 23.17 & \cellcolor[rgb]{0.76, 0.76, 0.76} 0.6305 & \cellcolor[rgb]{0.85, 0.85, 0.85} 0.4794 & \cellcolor[rgb]{0.79, 0.79, 0.79} 39.76\\ 
  &   &  A29   & \cellcolor[rgb]{0.93, 0.93, 0.93} 17.12 & \cellcolor[rgb]{0.98, 0.98, 0.98} 5.88 & \cellcolor[rgb]{1.00, 1.00, 1.00} 1.19 & \cellcolor[rgb]{0.87, 0.87, 0.87} 0.3988 & \cellcolor[rgb]{0.98, 0.98, 0.98} 0.0821 & \cellcolor[rgb]{0.85, 0.85, 0.85} 30.72\\ 
\cmidrule{3-9}
  & \multirow{7}{*}{\rotatebox{90}{Adversarial}}
    &   A18   & \cellcolor[rgb]{0.89, 0.89, 0.89} 21.96 & \cellcolor[rgb]{0.71, 0.71, 0.71} 53.24 & \cellcolor[rgb]{0.69, 0.69, 0.69} 52.24 & \cellcolor[rgb]{0.81, 0.81, 0.81} 0.5279 & \cellcolor[rgb]{0.88, 0.88, 0.88} 0.4029 & \cellcolor[rgb]{0.99, 0.99, 0.99} 1.53\\ 
  &   &  A20   & \cellcolor[rgb]{0.85, 0.85, 0.85} 27.21 & \cellcolor[rgb]{0.75, 0.75, 0.75} 48.54 & \cellcolor[rgb]{0.86, 0.86, 0.86} 30.72 & \cellcolor[rgb]{0.75, 0.75, 0.75} 0.6486 & \cellcolor[rgb]{0.63, 0.63, 0.63} 0.8789 & \cellcolor[rgb]{1.00, 1.00, 1.00} 0.09\\ 
  &   &  A23   & \cellcolor[rgb]{0.95, 0.95, 0.95} 13.33 & \cellcolor[rgb]{0.90, 0.90, 0.90} 25.08 & \cellcolor[rgb]{0.86, 0.86, 0.86} 30.02 & \cellcolor[rgb]{0.90, 0.90, 0.90} 0.3236 & \cellcolor[rgb]{0.94, 0.94, 0.94} 0.2321 & \cellcolor[rgb]{0.99, 0.99, 0.99} 2.47\\ 
  &   &  A27   & \cellcolor[rgb]{0.88, 0.88, 0.88} 23.42 & \cellcolor[rgb]{0.82, 0.82, 0.82} 37.94 & \cellcolor[rgb]{0.90, 0.90, 0.90} 22.91 & \cellcolor[rgb]{0.79, 0.79, 0.79} 0.5668 & \cellcolor[rgb]{0.94, 0.94, 0.94} 0.2553 & \cellcolor[rgb]{1.00, 1.00, 1.00} 0.06\\ 
  &   &  A30   & \cellcolor[rgb]{0.80, 0.80, 0.80} 33.35 & \cellcolor[rgb]{0.80, 0.80, 0.80} 41.82 & \cellcolor[rgb]{0.76, 0.76, 0.76} 43.84 & \cellcolor[rgb]{0.65, 0.65, 0.65} 0.8034 & \cellcolor[rgb]{0.89, 0.89, 0.89} 0.3772 & \cellcolor[rgb]{1.00, 1.00, 1.00} 0.33\\ 
  &   &  A31   & \cellcolor[rgb]{0.84, 0.84, 0.84} 28.39 & \cellcolor[rgb]{0.86, 0.86, 0.86} 32.39 & \cellcolor[rgb]{0.89, 0.89, 0.89} 24.12 & \cellcolor[rgb]{0.73, 0.73, 0.73} 0.6840 & \cellcolor[rgb]{0.92, 0.92, 0.92} 0.2901 & \cellcolor[rgb]{1.00, 1.00, 1.00} 0.62\\ 
  &   &  A32   & \cellcolor[rgb]{0.93, 0.93, 0.93} 17.02 & \cellcolor[rgb]{0.87, 0.87, 0.87} 29.58 & \cellcolor[rgb]{0.94, 0.94, 0.94} 15.98 & \cellcolor[rgb]{0.87, 0.87, 0.87} 0.4103 & \cellcolor[rgb]{0.94, 0.94, 0.94} 0.2457 & \cellcolor[rgb]{1.00, 1.00, 1.00} 0.03\\ 
  \cmidrule{3-9}
  &   & All attacks & \cellcolor[rgb]{0.88, 0.88, 0.88} 23.13 & \cellcolor[rgb]{0.86, 0.86, 0.86} 32.33 & \cellcolor[rgb]{0.89, 0.89, 0.89} 25.28 & \cellcolor[rgb]{0.81, 0.81, 0.81} 0.5240 & \cellcolor[rgb]{0.91, 0.91, 0.91} 0.3296 & \cellcolor[rgb]{0.94, 0.94, 0.94} 14.89\\ 
  \midrule
\multirow{18}{*}{\rotatebox{90}{Evaluation set \textbf{with} encoding/compression}}
  & & Non-target & \cellcolor[rgb]{0.98, 0.98, 0.98} 6.00 & \\ 
\cmidrule(l){3-9}
  & \multirow{9}{*}{\rotatebox{90}{TTS/VC }}
  &   A17   & \cellcolor[rgb]{0.69, 0.69, 0.69} 44.07 & \cellcolor[rgb]{0.90, 0.90, 0.90} 24.66 & \cellcolor[rgb]{0.92, 0.92, 0.92} 19.18 & \cellcolor[rgb]{0.59, 0.59, 0.59} 0.8863 & \cellcolor[rgb]{0.76, 0.76, 0.76} 0.6813 & \cellcolor[rgb]{0.72, 0.72, 0.72} 48.69\\ 
  &   &  A19   & \cellcolor[rgb]{0.64, 0.64, 0.64} 48.17 & \cellcolor[rgb]{0.62, 0.62, 0.62} 62.85 & \cellcolor[rgb]{0.62, 0.62, 0.62} 59.32 & \cellcolor[rgb]{0.59, 0.59, 0.59} 0.8924 & \cellcolor[rgb]{0.59, 0.59, 0.59} 0.9491 & \cellcolor[rgb]{0.94, 0.94, 0.94} 15.58\\ 
  &   &  A21   & \cellcolor[rgb]{0.82, 0.82, 0.82} 30.87 & \cellcolor[rgb]{0.88, 0.88, 0.88} 28.11 & \cellcolor[rgb]{0.92, 0.92, 0.92} 20.46 & \cellcolor[rgb]{0.72, 0.72, 0.72} 0.7070 & \cellcolor[rgb]{0.85, 0.85, 0.85} 0.4793 & \cellcolor[rgb]{0.76, 0.76, 0.76} 42.98\\ 
  &   &  A22   & \cellcolor[rgb]{0.86, 0.86, 0.86} 26.82 & \cellcolor[rgb]{0.89, 0.89, 0.89} 26.12 & \cellcolor[rgb]{0.91, 0.91, 0.91} 20.81 & \cellcolor[rgb]{0.77, 0.77, 0.77} 0.6063 & \cellcolor[rgb]{0.82, 0.82, 0.82} 0.5466 & \cellcolor[rgb]{0.82, 0.82, 0.82} 35.27\\ 
  &   &  A24   & \cellcolor[rgb]{0.90, 0.90, 0.90} 21.15 & \cellcolor[rgb]{0.89, 0.89, 0.89} 25.96 & \cellcolor[rgb]{0.94, 0.94, 0.94} 16.41 & \cellcolor[rgb]{0.84, 0.84, 0.84} 0.4790 & \cellcolor[rgb]{0.87, 0.87, 0.87} 0.4352 & \cellcolor[rgb]{0.80, 0.80, 0.80} 38.37\\ 
  &   &  A25   & \cellcolor[rgb]{0.96, 0.96, 0.96} 10.79 & \cellcolor[rgb]{0.86, 0.86, 0.86} 32.23 & \cellcolor[rgb]{0.89, 0.89, 0.89} 24.77 & \cellcolor[rgb]{0.93, 0.93, 0.93} 0.2561 & \cellcolor[rgb]{0.86, 0.86, 0.86} 0.4592 & \cellcolor[rgb]{0.96, 0.96, 0.96} 10.97\\ 
  &   &  A26   & \cellcolor[rgb]{0.93, 0.93, 0.93} 15.97 & \cellcolor[rgb]{0.78, 0.78, 0.78} 43.22 & \cellcolor[rgb]{0.83, 0.83, 0.83} 35.17 & \cellcolor[rgb]{0.88, 0.88, 0.88} 0.3842 & \cellcolor[rgb]{0.79, 0.79, 0.79} 0.6104 & \cellcolor[rgb]{0.88, 0.88, 0.88} 25.14\\ 
  &   &  A28   & \cellcolor[rgb]{0.75, 0.75, 0.75} 39.20 & \cellcolor[rgb]{0.80, 0.80, 0.80} 41.40 & \cellcolor[rgb]{0.83, 0.83, 0.83} 35.05 & \cellcolor[rgb]{0.61, 0.61, 0.61} 0.8553 & \cellcolor[rgb]{0.64, 0.64, 0.64} 0.8738 & \cellcolor[rgb]{0.59, 0.59, 0.59} 61.64\\ 
  &   &  A29   & \cellcolor[rgb]{0.80, 0.80, 0.80} 33.07 & \cellcolor[rgb]{0.92, 0.92, 0.92} 20.13 & \cellcolor[rgb]{0.96, 0.96, 0.96} 11.20 & \cellcolor[rgb]{0.69, 0.69, 0.69} 0.7507 & \cellcolor[rgb]{0.87, 0.87, 0.87} 0.4277 & \cellcolor[rgb]{0.65, 0.65, 0.65} 55.39\\ 
\cmidrule{3-9}
  & \multirow{7}{*}{\rotatebox{90}{Adversarial}}
    &  A18   & \cellcolor[rgb]{0.82, 0.82, 0.82} 31.05 & \cellcolor[rgb]{0.65, 0.65, 0.65} 59.26 & \cellcolor[rgb]{0.72, 0.72, 0.72} 49.08 & \cellcolor[rgb]{0.73, 0.73, 0.73} 0.6870 & \cellcolor[rgb]{0.72, 0.72, 0.72} 0.7415 & \cellcolor[rgb]{0.96, 0.96, 0.96} 11.17\\ 
  &   &  A20   & \cellcolor[rgb]{0.83, 0.83, 0.83} 30.51 & \cellcolor[rgb]{0.76, 0.76, 0.76} 46.38 & \cellcolor[rgb]{0.83, 0.83, 0.83} 34.30 & \cellcolor[rgb]{0.75, 0.75, 0.75} 0.6583 & \cellcolor[rgb]{0.64, 0.64, 0.64} 0.8652 & \cellcolor[rgb]{0.99, 0.99, 0.99} 1.62\\ 
  &   &  A23   & \cellcolor[rgb]{0.86, 0.86, 0.86} 26.07 & \cellcolor[rgb]{0.86, 0.86, 0.86} 31.81 & \cellcolor[rgb]{0.84, 0.84, 0.84} 32.63 & \cellcolor[rgb]{0.78, 0.78, 0.78} 0.5990 & \cellcolor[rgb]{0.82, 0.82, 0.82} 0.5353 & \cellcolor[rgb]{0.94, 0.94, 0.94} 16.04\\ 
  &   &  A27   & \cellcolor[rgb]{0.84, 0.84, 0.84} 28.21 & \cellcolor[rgb]{0.81, 0.81, 0.81} 39.10 & \cellcolor[rgb]{0.86, 0.86, 0.86} 29.20 & \cellcolor[rgb]{0.77, 0.77, 0.77} 0.6136 & \cellcolor[rgb]{0.80, 0.80, 0.80} 0.5965 & \cellcolor[rgb]{1.00, 1.00, 1.00} 1.38\\ 
  &   &  A30   & \cellcolor[rgb]{0.79, 0.79, 0.79} 34.66 & \cellcolor[rgb]{0.79, 0.79, 0.79} 42.29 & \cellcolor[rgb]{0.77, 0.77, 0.77} 42.32 & \cellcolor[rgb]{0.70, 0.70, 0.70} 0.7307 & \cellcolor[rgb]{0.76, 0.76, 0.76} 0.6753 & \cellcolor[rgb]{0.99, 0.99, 0.99} 4.12\\ 
  &   &  A31   & \cellcolor[rgb]{0.83, 0.83, 0.83} 30.23 & \cellcolor[rgb]{0.85, 0.85, 0.85} 33.41 & \cellcolor[rgb]{0.87, 0.87, 0.87} 28.59 & \cellcolor[rgb]{0.75, 0.75, 0.75} 0.6469 & \cellcolor[rgb]{0.81, 0.81, 0.81} 0.5685 & \cellcolor[rgb]{0.98, 0.98, 0.98} 6.64\\ 
  &   &  A32   & \cellcolor[rgb]{0.88, 0.88, 0.88} 23.87 & \cellcolor[rgb]{0.87, 0.87, 0.87} 29.73 & \cellcolor[rgb]{0.92, 0.92, 0.92} 20.53 & \cellcolor[rgb]{0.81, 0.81, 0.81} 0.5300 & \cellcolor[rgb]{0.81, 0.81, 0.81} 0.5666 & \cellcolor[rgb]{1.00, 1.00, 1.00} 1.01\\ 
  \cmidrule{3-9}
  &   & All attacks & \cellcolor[rgb]{0.83, 0.83, 0.83} 30.18 & \cellcolor[rgb]{0.83, 0.83, 0.83} 37.11 & \cellcolor[rgb]{0.86, 0.86, 0.86} 30.61 & \cellcolor[rgb]{0.75, 0.75, 0.75} 0.6500 & \cellcolor[rgb]{0.78, 0.78, 0.78} 0.6284 & \cellcolor[rgb]{0.86, 0.86, 0.86} 29.74\\ 
\bottomrule
\end{tabular}
\label{tab:baseline_codec}
\end{table}

\clearpage
\newpage
\bibliographystyle{elsarticle-num}
\bibliography{refs}
\end{document}